\documentclass[12pt,indent]{article}
\usepackage{a4,latexsym,amsmath,amssymb}
\usepackage[latin1]{inputenc}
\usepackage{float}

\usepackage{booktabs,caption}
\usepackage[flushleft]{threeparttable}
\usepackage{rotating}

\usepackage{natbib}
\usepackage{graphics}
\usepackage[english]{babel}
\usepackage{tabularx}
\usepackage{lscape}
\usepackage{multirow}
\usepackage{rotating}
\usepackage{dsfont}
\usepackage{caption}
\usepackage{subcaption}
\usepackage{bbm}
\usepackage[table]{xcolor}
\usepackage{booktabs}
\usepackage{calc}
\usepackage{ifthen}
\usepackage{url}
\usepackage{colortbl}      
\usepackage{tikz}

\usepackage{blindtext}
\usepackage{scrextend}
\usepackage{mathtools}
\usepackage{verbatim}

\usetikzlibrary{shapes,snakes,matrix}

\newenvironment{tabularsmall}
{ \footnotesize \sffamily \tabular } {
\endtabular
\normalfont }

\newcommand{\E}{\operatorname{E}}      

\newcommand{\NV}{\operatorname{N}}

\newcommand{\deltab}{\boldsymbol{\delta}}

\newcommand{\pib}{{\boldsymbol{\pi}}}

\newcommand{\yb}{\boldsymbol{y}}

\newcommand{\blanco}[1]{}

\def\d{\displaystyle}

\usepackage{natbib}

\usepackage{geometry}
\usepackage{setspace}
\usepackage{tikz}
\usepackage[]{graphicx}
\usepackage[]{color}
\makeatletter
\def\maxwidth{ %
  \ifdim\Gin@nat@width>\linewidth
    \linewidth
  \else
    \Gin@nat@width
  \fi
}
\makeatother

\definecolor{fgcolor}{rgb}{0.345, 0.345, 0.345}

\usepackage{framed}
\makeatletter
 {\par\unskip\endMakeFramed%
 \at@end@of@kframe}
\makeatother

\definecolor{shadecolor}{rgb}{.97, .97, .97}
\definecolor{messagecolor}{rgb}{0, 0, 0}
\definecolor{warningcolor}{rgb}{1, 0, 1}
\definecolor{errorcolor}{rgb}{1, 0, 0}

\usepackage{alltt}
\IfFileExists{upquote.sty}{\usepackage{upquote}}{}

\begin{document}
\bibliographystyle{chicago}
\sloppy

\makeatletter
\renewcommand{\section}{\@startsection{section}{1}{\z@}%
        {-3.5ex \@plus -1ex \@minus -.2ex}%
        {1.5ex \@plus.2ex}%
        {\reset@font\large\sffamily}}
\renewcommand{\subsection}{\@startsection{subsection}{1}{\z@}%
        {-3.25ex \@plus -1ex \@minus -.2ex}%
        {1.1ex \@plus.2ex}%
        {\reset@font\normalsize\sffamily\flushleft}}
\renewcommand{\subsubsection}{\@startsection{subsubsection}{1}{\z@}%
        {-3.25ex \@plus -1ex \@minus -.2ex}%
        {1.1ex \@plus.2ex}%
        {\reset@font\normalsize\sffamily\flushleft}}
\makeatother



\newsavebox{\tempbox}
\newlength{\linelength}
\setlength{\linelength}{\linewidth-10mm} \makeatletter
\renewcommand{\@makecaption}[2]
{
  \renewcommand{\baselinestretch}{1.1} \normalsize\small
  \vspace{5mm}
  \sbox{\tempbox}{#1: #2}
  \ifthenelse{\lengthtest{\wd\tempbox>\linelength}}
  {\noindent\hspace*{4mm}\parbox{\linewidth-10mm}{\sc#1: \sl#2\par}}
  {\begin{center}\sc#1: \sl#2\par\end{center}}
}



\def\R{\mathchoice{ \hbox{${\rm I}\!{\rm R}$} }
                   { \hbox{${\rm I}\!{\rm R}$} }
                   { \hbox{$ \scriptstyle  {\rm I}\!{\rm R}$} }
                   { \hbox{$ \scriptscriptstyle  {\rm I}\!{\rm R}$} }  }

\def\N{\mathchoice{ \hbox{${\rm I}\!{\rm N}$} }
                   { \hbox{${\rm I}\!{\rm N}$} }
                   { \hbox{$ \scriptstyle  {\rm I}\!{\rm N}$} }
                   { \hbox{$ \scriptscriptstyle  {\rm I}\!{\rm N}$} }  }

\def\d{\displaystyle}\def\d{\displaystyle}

\title{Invariance of Comparisons: Separation of Item and Person Parameters beyond  Rasch Models }

  \author{Gerhard Tutz \\{\small Ludwig-Maximilians-Universit\"{a}t M\"{u}nchen}\\{\small Akademiestra{\ss}e 1, 80799 M\"{u}nchen}}

\maketitle
\begin{abstract} 
\noindent
The Rasch model is the most prominent member of the class of latent trait models that are in common use. The main reason is that it can be considered as a measurement model that allows to separate person and item parameters, a feature that is referred to as invariance of comparisons or specific objectivity. It is shown that the property is not an exclusive trait of Rasch type models but is also found in alternative latent trait models. It is distinguished between separability in the theoretical measurement model and empirical separability with empirical separability meaning that parameters can be estimated without reference to the other group of parameters. A new type of pairwise estimator with this property is proposed that can be used also in alternative models. Separability is considered in binary   
models as well as in polytomous models.
\end{abstract}

\noindent{\bf Keywords:} Rasch model; specific objectivity; invariance of comparisons, latent trait models

\section{Introduction}
The binary Rasch model \citep{rasch1961general} is one of the cornerstones of modern item response theory and has been extended to a whole family 
of models, see for example, \citet{rasch1960,Masters:82,Andrichh2016,fischer2012rasch,von2007multivariate,von2016rasch}.  The main advantage of the Rasch model is that it allows measurement of latent traits to be independent of the measurement instrument, which is considered an advantage  over other models as, for example, the normal-ogive model considered by \citet{lord1952theory}.

With reference to stimuli rather than items \citet{rasch1961general} formulated the requirements for comparing individuals and comparing stimuli by stating:
\medskip
\begin{addmargin}[0.5 cm]{-0.0 cm}
\small
The comparison between two stimuli should be independent of which particular individuals were instrumental for the comparison; and it should also be independent
of which other stimuli within the considered class were or might also have been
compared.

Symmetrically, a comparison between two individuals should be independent of
which particular stimuli within the class considered were instrumental for the comparison;
and it should also be independent of which other individuals were also
compared, on the same or on some other occasion  \citep{rasch1961general}, p.331.\\
\end{addmargin}

The strength of the Rasch model is that it allows for comparisons of item difficulties (person abilities) that are independent of the selection of persons (items). Rasch used the term \textit{specific objectivity} for measurements that allow for the comparison of subjects or objects without reference to the instrument (within a given well-defined frame of reference), and considered it as general scientific principle to obtain generalizable  measurements. The property has also be referred to as \textit{parameter separability} or \textit{invariance of comparisons}.  

Although there are stringent mathematical arguments why the Rasch model, which uses logistic item characteristic curves, allows for the separation of parameters it is hardly convincing that the  normal-ogive model, which in applications typically shows very similar results and fits, should not allow for invariant comparisons. It is demonstrated that parameter separability is possible in a much wider class of models, although not by means of conditional likelihood as in the Rasch model.
When investigating separability it is distinguished between separability as a property of the probabilistic measurement model and empirical separability as an estimation problem.  While the former is easily seen to hold for more general models than the Rasch model separate estimation of  parameters is less straightforward.

In Section \ref{sec:invR} invariance in the Rasch model is briefly considered. In Section \ref{sec:compbin} invariance is investigated for the wider class of monotone homogeneity models and an estimator is derived that separates parameters. In Section \ref{sec:further} further properties of the estimator are considered and a motivation as a smoothing method is given. It is also investigated what independence of parameter estimates means, which is often misunderstood. Section \ref{sec:poly} it devoted to polytomous models, in particular separability in the graded response model, which is not a member of the Rasch family, is considered.

\section{Invariance of Comparisons in the Binary Rasch model}\label{sec:invR}

Let  $Y_{pi} \in \{0,1\}$ denote the response of person $p$ on item $i$. The binary Rasch model can be given by
\begin{equation}\label{eq:BinRasch}
P(Y_{pi}=1|\theta_p,\delta_{i})=\frac{\exp(\theta_p-\delta_{i})}{1+\exp(\theta_p-\delta_{i})},  i=1,\dots,I, p=1,\dots, P,
\end{equation} 
where  $\theta_p$ is the ability of person $p$, and  $\delta_{i}$ is the difficulty of item $i$.
A key property concerning parameter separability can be derived by  considering odds. The odds of a response $Y_{pi} =1$ are given by 
\[
\gamma_{pi}=\frac{P(Y_{pi}=1|\theta_p,\delta_{i})}{P(Y_{pi}=0|\theta_p,\delta_{i})}= \exp(\theta_p-\delta_{i}).
\]
Then the odds ratio for two persons $p_1, p_2$ is  given by
\begin{align}\label{eq:Raschspecobj}
\frac{\gamma_{p_1 i}}{\gamma_{p_2 i}}=\frac{\exp(\theta_{p_1})}{\exp(\theta_{p_2})}=\frac{\theta_{p_1}^*}{\theta_{p_2}^*},
\end{align}
where $\theta_{p}^*=\exp(\theta_{p})$. That means comparison of persons can be carried out independently of the items involved, person parameters can be separated from item parameters. \citet{rasch1961general} preferred the parameterization $\theta_{p}^*$, which yields the proportion $\theta_{p_1}^*/{\theta_{p_2}^*}$ when comparing odds of two items. Here, we will mostly use the parameterization $\theta_{p}$, which is linked to differences of parameters rather than the proportion since $\exp(\theta_{p_1})/\exp(\theta_{p_2})= \exp(\theta_{p_1}-\theta_{p_2})$

Since the model is symmetric in the parameters one obtains a similar result for the  comparison of two items $i_1, i_2$,
\begin{align}\label{eq:Raschsit}
\frac{\gamma_{p i_1}}{\gamma_{p i_2}}=\frac{\exp(-\delta_{i_1})}{\exp(-\delta_{i_2})}=\frac{\delta_{i_1}^*}{\delta_{i_2}^*},
\end{align}
where $\delta_{i}^*=\exp(-\delta_{i})$, which does not depend on the person. Thus, comparisons of items can be carried out independent of the persons involved.

Equations (\ref{eq:Raschspecobj}) and (\ref{eq:Raschsit}) show that parameters can be separated by using odds, however odds that are not directly observed.
In general, in probabilistic models inference tools are needed that approximate unobservable terms. In  Rasch models a possible path to \textit{empirical} separation of parameters is based on exploiting that the total scores are sufficient statistics. Let $S_p= Y_{p+}=\sum_{i=1}^I y_{pi}$ denote the number of  items solved by person $p$,
that is, the total score of person $p$. Then one can derive that 
\begin{align}\label{eq:cestit}
P( Y_{p1}=y_{p1},\dots,Y_{pI}=y_{pI}|S_p=s) = \frac{e^{-\sum_{i=1}^I y_{pi}\delta_i}}{\gamma_{s}(\deltab)},
\end{align}
where the functions $
\gamma_{s}(\deltab)=\sum_{a_i \in \{0,1\},\sum_{i} a_i=s}  e^{-\sum_{i}  a_i\delta_i} 
$
are the so-called symmetric functions of order $s$, depending on $\deltab^T=(\delta_1,\dots,\delta_I)$ only.
Since the conditional probability given in (\ref{eq:cestit}) is a function of item parameters only it can be used to estimate item parameters irrespective of the persons involved by maximizing the conditional likelihood $L_c (\deltab) = \prod_{p} P( Y_{p1}=y_{p1},\dots,Y_{pI}=y_{pI}|\sum_{i=1}^I  Y_{pi}= y_{p+})$.
For the asymptotic distribution and necessary and sufficient conditions for the existence of estimates see  \citet{Andersen:77}, \citet{pfanzagl1994item} and
\citet{fischer1981existence}.

\section{ Comparisons in Binary Latent Trait Models}\label{sec:compbin}

Let us consider the more general class of monotone homogeneity models (MH models), which comprises models of the form
\begin{equation}\label{eq:BinRasch}
\pi_{pi}(\theta_p,\delta_{i})=P(Y_{pi}=1|\theta_p,\delta_{i})=F(\theta_p-\delta_{i}),
\end{equation} 
where $F(.)$ is a strictly monotone distribution function also called response function.
The models contain one  parameter per person, $\theta_p$, and one parameter per item, $\delta_{i}$.
They are homogeneous since the item characteristic functions all have the same form. They are  monotone since the probability of success increases monotonically with increasing person parameter. The binary Rasch model is contained as the special case where $F(.)$ is the logistic function $F(\eta)=\exp(\eta)/(1+\exp(\eta))$.

For a MH model one obtains a form of separability of parameters when considering two items $i_1,i_2$. From (\ref{eq:BinRasch}) one can immediately derive
\begin{equation}\label{eq:MHso}
F^{-1}(\pi_{pi}(\theta_p,\delta_{i_1}))-F^{-1}(\pi_{pi}(\theta_p,\delta_{i_2}))= \delta_{i_2}-\delta_{i_1},
\end{equation}
which is independent of $\theta_p$. Thus a transformation of the involved probabilities yields a function that does not contain person parameters, and which can be used to compare item parameters. Since the function $F^{-1}(.)$ is the quantile function  differences of item parameters reflect the differences of quantiles 
of the distribution function $F(.)$, which do not depend on the person parameters.
In a similar way persons can be compared without reference to item parameters.

\subsection{Invariance in Terms of the Model}

The  separability property (\ref{eq:MHso}) can be formulated in a more general way. 
For a  uni-dimensional latent trait model with response probabilities $\pi_{pi}=P(Y_{pi}=1)$ for person $p$ and item $i$ 
\textit{separability of item parameters from person parameters}
(\textit{invariance of comparison of item parameters}, \textit{specific objectivity for the comparison of items}) holds if  a parameterization $(\theta_p,\delta_{i})$ and a transformation function $C_{\text{it}}$ called comparator   exist such that $C_{\text{it}}(\pi_{pi_1},\pi_{p i_2})$ is equal to the difference of item parameters  for all items $i_1,i_2$.  
For the parameterization that is assumed to exist one has
\begin{align}\label{eq:SO}
C_{\text{it}}(\pi_{pi_1}(\theta_p,\delta_{i_1}),\pi_{p i_2}(\theta_p,\delta_{i_1})) = \delta_{pi_1}- \delta_{pi_2}. 
\end{align}
The definition of separability only assumes that a parameterization exists in order to make the definition independent of the specific parameterization that is used in the formulation of the model. For example, in the  Rasch model the parameterization $\theta_p,\delta_{i}$ provides such a parameterization but the parameterization $\theta_p^*,\delta_{i}^*$ does not. \citet{fischer1995derivations} used a similar equation when investigating invariance in the Rasch model. He assumed for a fixed parameterization $C(\rho_{pi_1}(\theta_p,\delta_{i_1}),\rho_{p i_2}(\theta_p,\delta_{i_1})) = V(\delta_{pi_1}, \delta_{pi_2})$, where $V(.)$ is an additional function and $\rho_{pi}$ a ``reaction parameter''. The definition used here considers the response probability as reaction parameter, which seems quite natural since it  definitely determines the response. A general function $V(.)$ seems not necessary if one does not consider a specific parameterization but assumes the existence of a parameterization. In addition, an unspecified function  $V(.)$ could be very difficult while differences are easy to handle and suffice for the models considered here. Although a similar definition can be given concerning the invariance of comparison of person parameters
in the following we focus on the invariance of comparison of item parameters.

It follows from (\ref{eq:MHso}) that in any monotone homogeneity model the comparison of item parameters is invariant  and does not depend on the person parameter. 
The Rasch model is just a special case but by far not the only model. It is to be emphasized that the invariance property considered here is a property of the probabilistic measurement model
but is not directly observable. However, it can be considered as the essential property that is needed to also obtain empirical invariance, that is, invariance    
referring to estimation.

\subsection{Empirical Invariance}

In the Rasch model empirical separation of parameters is usually obtained by exploiting that the total scores are sufficient statistics. This works only since the item response curves are logistic, for any other response function total scores are not sufficient statistics. The deeper reason is that binary responses are members of the exponential family and logits are linked to the natural parameter in exponential families, see, for example, \citet{McCNel:89}.

Conditional maximum likelihood estimation with the conditioning on sufficient statistics is not an option if the response function is not the logistic function. 
Although in  monotone homogeneity models the function $F^{-1}(\pi_{pi}(\theta_p,\delta_{i_1}))-F^{-1}(\pi_{pi}(\theta_p,\delta_{i_2}))$ does not depend on person parameters it is not obvious how this property can be exploited in estimation since replacing $\pi_{pi}(\theta_p,\delta_{i})$ by observations, that is, by 0 or 1, yields $-\infty$ and $\infty$ when building $F^{-1}(Y_{pi})$.

In the following an estimation method is proposed that uses pseudo observations.
Let us consider the pseudo observations 
\[
Y_{pi}^{*}=Y_{pi}(1-2\gamma) + \gamma= \begin{cases}
\gamma &Y_{pi}=0 \\   
1-\gamma &Y_{pi}=1,\\
\end{cases}
\]
where $\gamma > 0$ is a fixed value. The pseudo observations approximate the original values, and in the extreme case are identical, $\lim_{\gamma\rightarrow 0}Y_{pi}^{*}=Y_{pi}$. Replacing probabilities by pseudo observations yields
\[
F^{-1}(Y_{pi}(1-2\gamma) + \gamma)= \begin{cases}
F^{-1}(\gamma)=:\gamma_0 &Y_{pi}=0 \\   
F^{-1}(1-\gamma)=:\gamma_1 &Y_{pi}=1.\\
\end{cases}
\]
Although $\lim_{\gamma \rightarrow 0}F^{-1}(\gamma)=-\infty$, $\lim_{\gamma \rightarrow 0}{F^{-1}(1-\gamma)}=\infty$, for values $\gamma>0$ one obtains finite values for $\gamma_0$ and $\gamma_1$.
The empirical analogue to  (\ref{eq:MHso}) when replacing probabilities by pseudo observations is
\begin{align*}
\hat\delta_{i_2}(p)-\hat\delta_{i_1}(p)&= F^{-1}(Y_{pi_1}(1-2\gamma) + \gamma)-F^{-1}(Y_{pi_2}(1-2\gamma) + \gamma)=\\
&=\begin{cases}
0 &Y_{pi_1}=Y_{pi_2} \\   
\gamma_1-\gamma_0 &Y_{pi_1}=1,Y_{pi_2}=0\\
\gamma_0-\gamma_1 &Y_{pi_1}=0,Y_{pi_2}=1\\
\end{cases}
\end{align*}
It is an empirical approximation of $F^{-1}(\pi_{pi}(\theta_p,\delta_{i_1}))-F^{-1}(\pi_{pi}(\theta_p,\delta_{i_2}))$, which is the theoretical difference for person $p$.
From this representation an estimate of the difference $\delta_{i_2 i_1}=\delta_{i_2}-\delta_{i_1}$ is derived by summing over the contributions of all persons, 
\begin{align*}
\hat\delta_{i_2 i_1}=\hat\delta_{i_2}-\hat\delta_{i_1}&=\{(\gamma_1-\gamma_0)n(1,0)+(\gamma_0-\gamma_1)n(0,1)\}/n(Y_{pi_1}\ne Y_{pi_2})\\
&=\gamma_{10}(n(1,0)-n(0,1))/n(Y_{pi_1}\ne Y_{pi_2})
\end{align*}
where $\gamma_{10}=\gamma_1-\gamma_0$, $n(1,0)$ is the number of persons with response $Y_{pi_1}=1,Y_{pi_2}=0$, $n(0,1)$ is the number of persons with response $Y_{pi_1}=0,Y_{pi_2}=1$,
and $n(Y_{pi_1}\ne Y_{pi_2})=n(1,0)+n(0,1)$. An alternative representation of the estimator  is 
\begin{align*}
\hat\delta_{i_2 i_1}&=\gamma_{10} \sum_{p=1}^P (Y_{pi_1}-Y_{pi_2})/n(Y_{pi_1}\ne Y_{pi_2}) = \gamma_{10}(Y_{+i_1}-Y_{+i_2})/n(Y_{pi_1}\ne Y_{pi_2}),
\end{align*} 
where $Y_{+i}=\sum_p Y_{pi}$ are the number of persons that solved item $i$. If $F(.)$ is a symmetric function, for example the normal distribution, one has $\gamma_0=-\gamma_1$, and the estimator simplifies to
$\hat\delta_{i_2}-\hat\delta_{i_1}= 2 \gamma_1(Y_{+i_1}-Y_{+i_2})/n(Y_{pi_1}\ne Y_{pi_2}).$

Since parameters are only defined up to an additive constant one can set $\delta_{1}=0$. That yields a simple estimator of $\delta_{i}$ by using
$\hat\delta_{i}= \gamma_{10}(Y_{+i}-Y_{+1})/n(Y_{pi}\ne Y_{p1})$. A disadvantage is that it uses only the item pairs $(1,2),(1,3),\dots,(1,I)$.
An estimator that uses all the differences and is called the \textit{pairwise separation estimator} is given by 
\begin{align*}
\hat\delta_{i}= \sum_{j=1}^I(\hat\delta_{ij}-\hat\delta_{1j})/I,
\end{align*}
where $\hat\delta_{11}:=0$. It is an average across all estimators that use the estimated differences to a fixed item with the  constraint $\delta_{1}=0$. 
This can be seen by considering a fixed anchor item $j$. Then $\hat\delta_{ij}$ is an estimator of $\delta_{i}-\delta_{j}$. It aims at estimating  the differences between item $i$ and item $j$. If one sets $\hat\delta_{i}=\hat\delta_{ij}$ for all $i$ implicitly the parameter $\hat\delta_{j}$ is set to zero. In order to set $\hat\delta_{1}=0$ one has to subtract $\hat\delta_{1j}$.

The estimate contains the scaling factor $\gamma_{10}$. While proportions $\hat\delta_{i}/\hat\delta_{j}$ do not depend on the scaling, the estimates $\hat\delta_{i}$ themselves do. Therefore, the scaling factor, which is determines   the definition of the pseudo observations, has to be chosen separately. 
A data-based approach to selecting the scale parameter is the following. 
For fixed $\gamma_{10}$ and the resulting item parameter estimates $\hat\delta_{i}(\gamma_{10})$ person parameters can be estimated by maximizing the log-likelihood function 
\begin{align*}
l_{\gamma_{10}}(\theta_p)= \sum_{i=1} ^I  Y_{pi} \log (\pi_{pi}(\theta_p,\hat\delta_{i}(\gamma_{10}))) + (1-Y_{pi})\log(1-\pi_{pi}(\theta_p,\hat\delta_{i}(\gamma_{10}))),
\end{align*}        
yielding $\hat\theta_p(\gamma_{10})$. Maximization is simple since it is a one-dimensional maximization problem. In a second step the goodness-of-fit of the resulting estimate $\pi_{pi}(\hat\theta_p(\gamma_{10}),\hat\delta_{i}(\gamma_{10}))$
is investigated by  considering loss functions that reflect the differences between observations and estimated probabilities, $L(Y_{pi},\pi_{pi}(\hat\theta_p(\gamma_{10},\hat\delta_{i}(\gamma_{10})))$. Candidates are the quadratic loss
$L_Q (Y_{pi}),\hat\pi_{pi}) = 2 (Y_{pi}-\hat\pi_{pi})^2$
and the Kullback-Leibler loss  
$L_{KL} (Y_{pi},\hat\pi_{pi}) = -(Y_{pi}\log(\hat\pi_{pi}) -(1-Y_{pi})\log(1-\hat\pi_{pi}))$, where
minimization of the latter  corresponds to maximum likelihood estimation.  
The final estimator is obtained by using that scaling parameter that minimizes 
\begin{align*}
\text{Loss}(\gamma_{10})= \sum_{i=1}^I \sum_{p=1}^P L(Y_{pi},\pi_{pi}(\hat\theta_p(\gamma_{10},\hat\delta_{i}(\gamma_{10}))) 
\end{align*} 
with respect to $\gamma_{10}$.

\begin{figure}[H]
\centering
\includegraphics[width=6.5 cm]{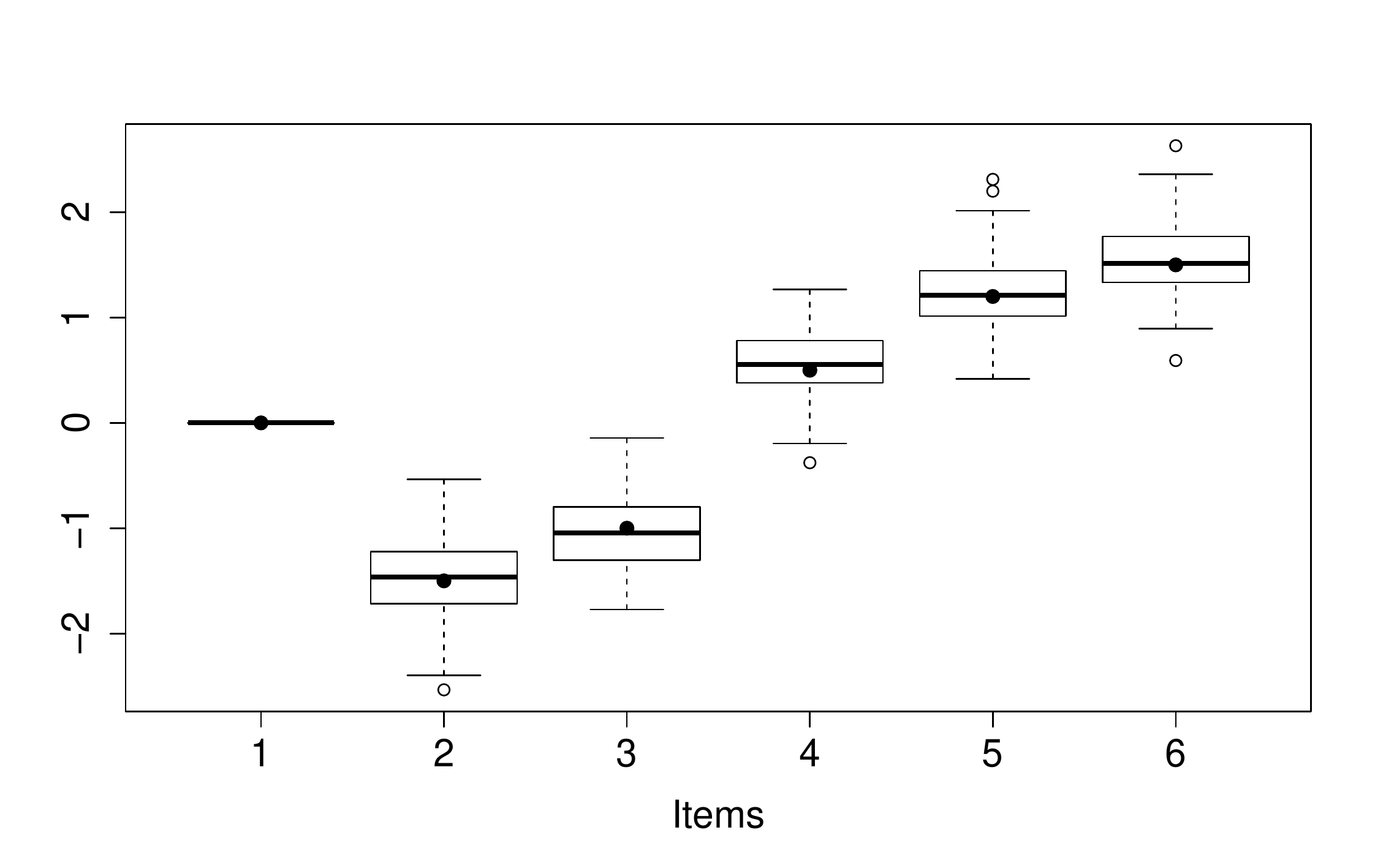}
\includegraphics[width=6.5 cm]{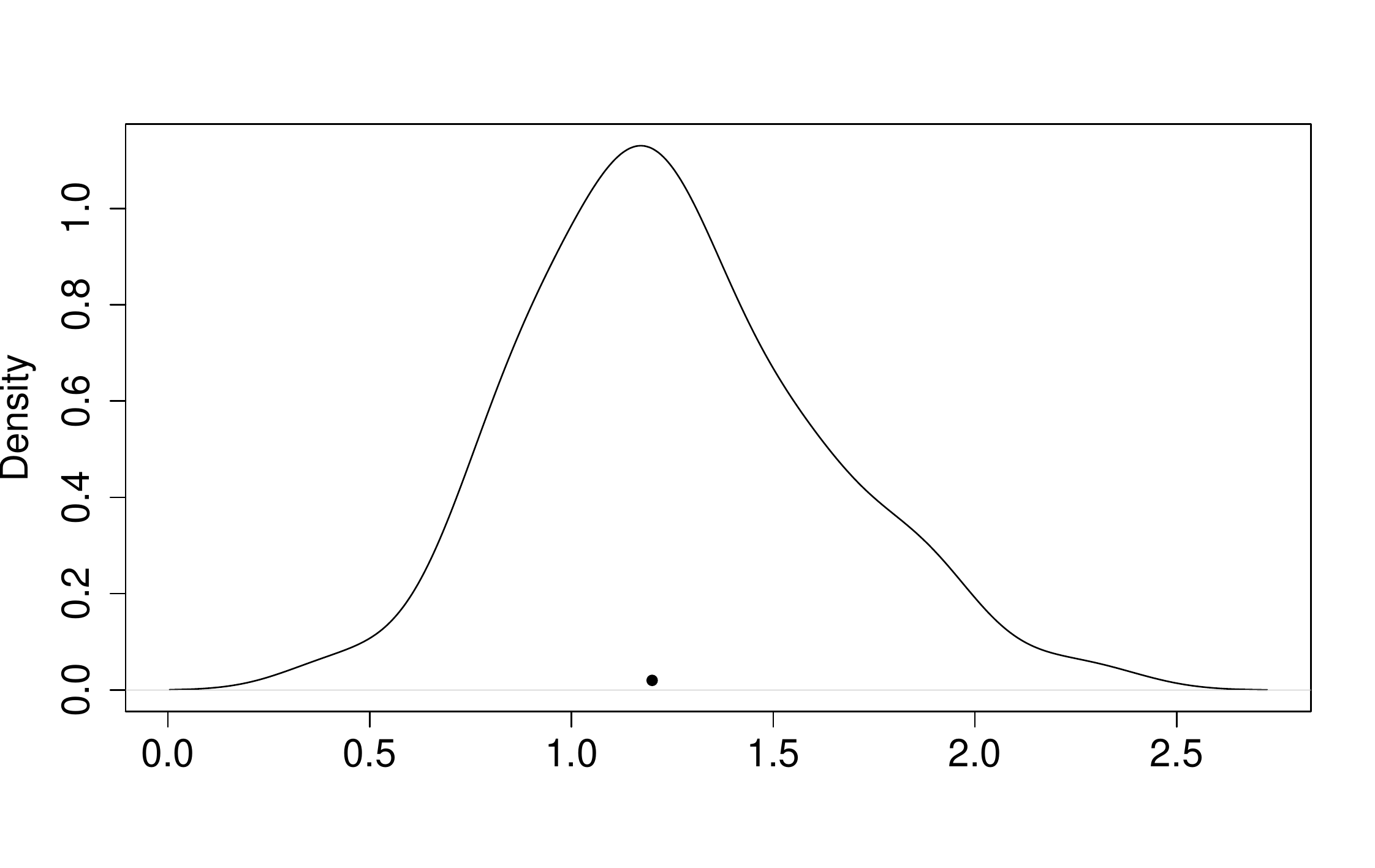}
\includegraphics[width=6.5 cm]{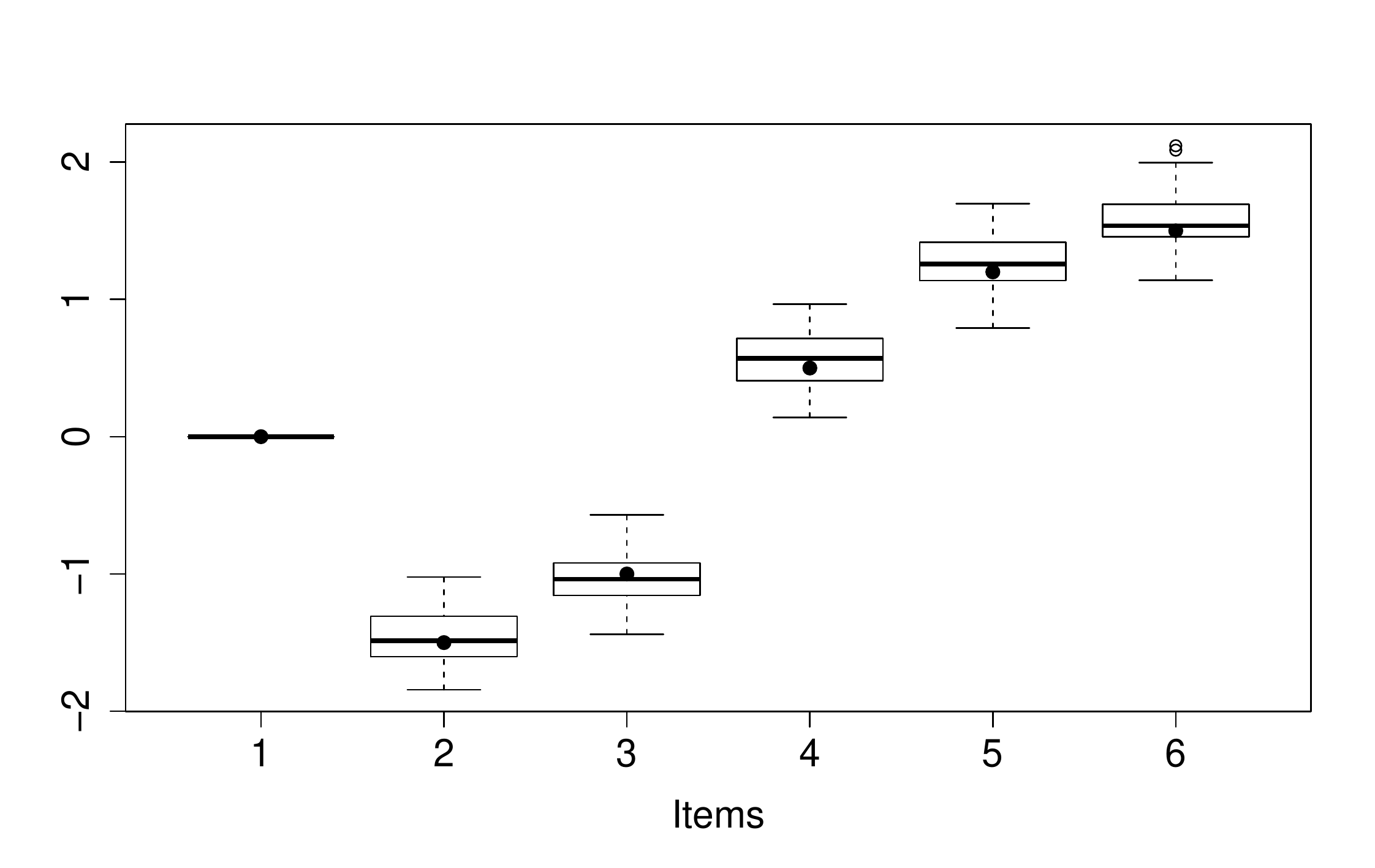}
\includegraphics[width=6.5 cm]{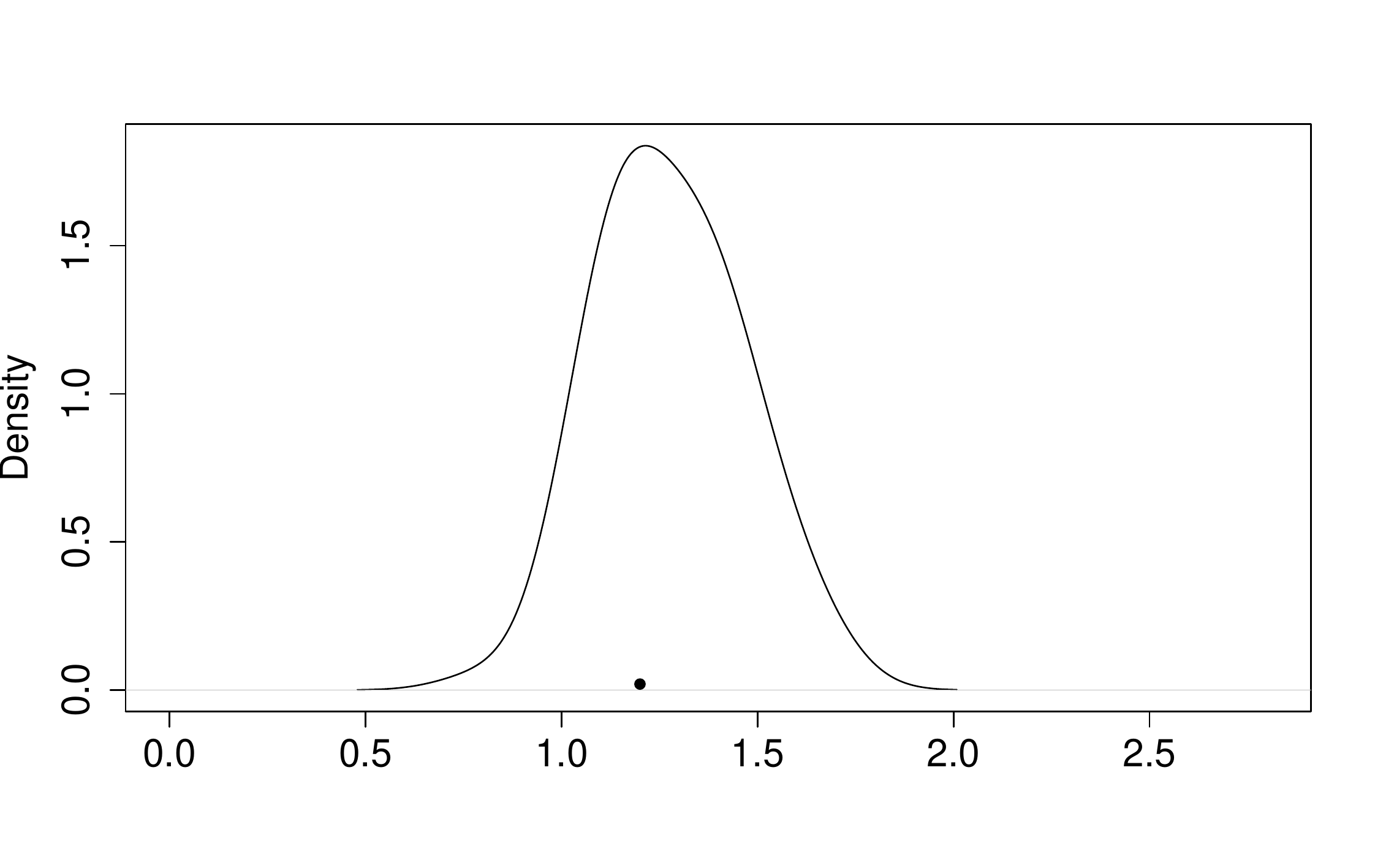}
\caption{Left: Box plots for estimates for six items with dots indicating the true values assuming a normal-ogive model; right: density of estimates for item 5. First row: (P=100), second row: $P=300$}.
\label{fig:NV1}
\end{figure}

\begin{figure}[H]
\centering
\includegraphics[width=6.5 cm]{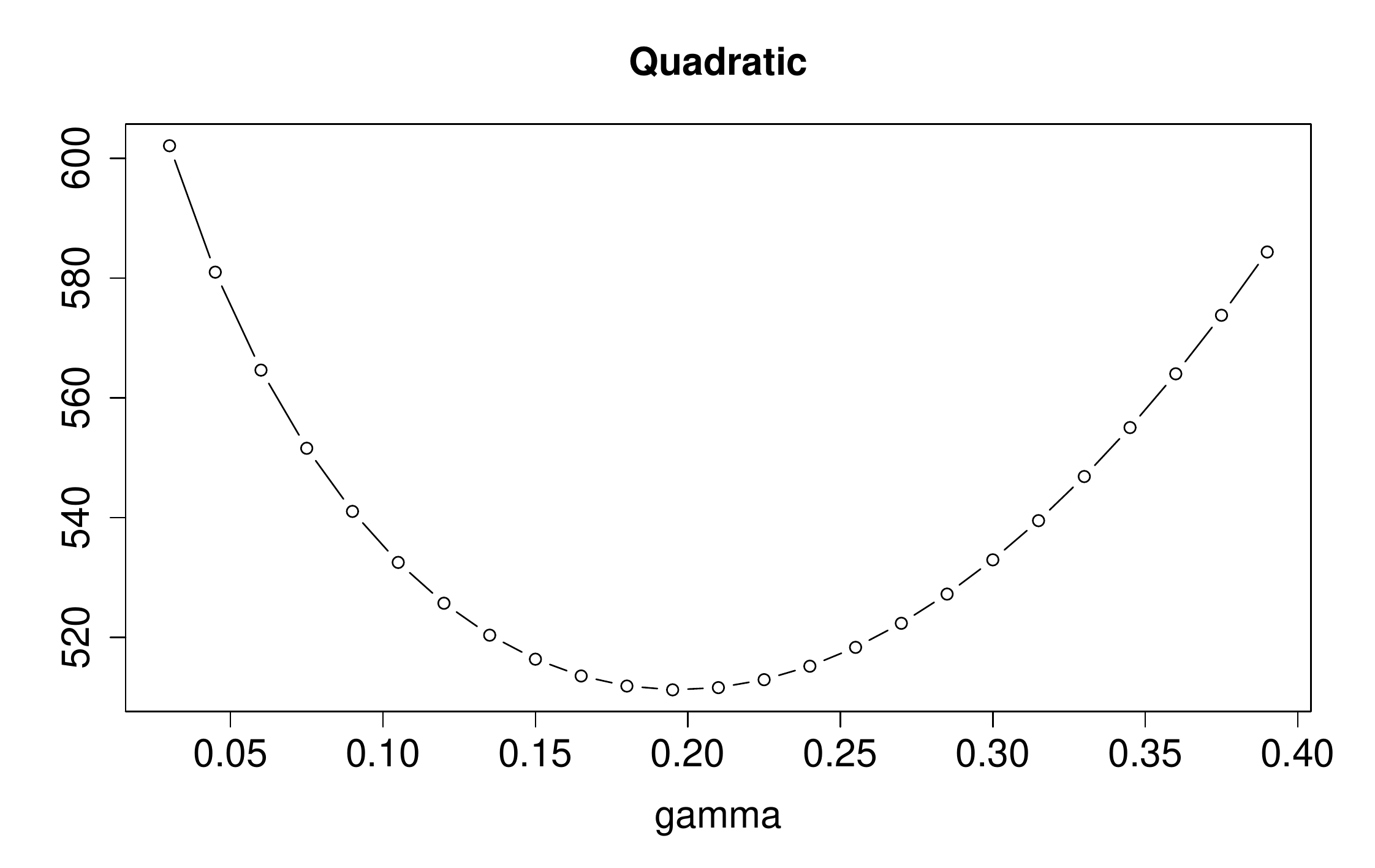}
\includegraphics[width=6.5 cm]{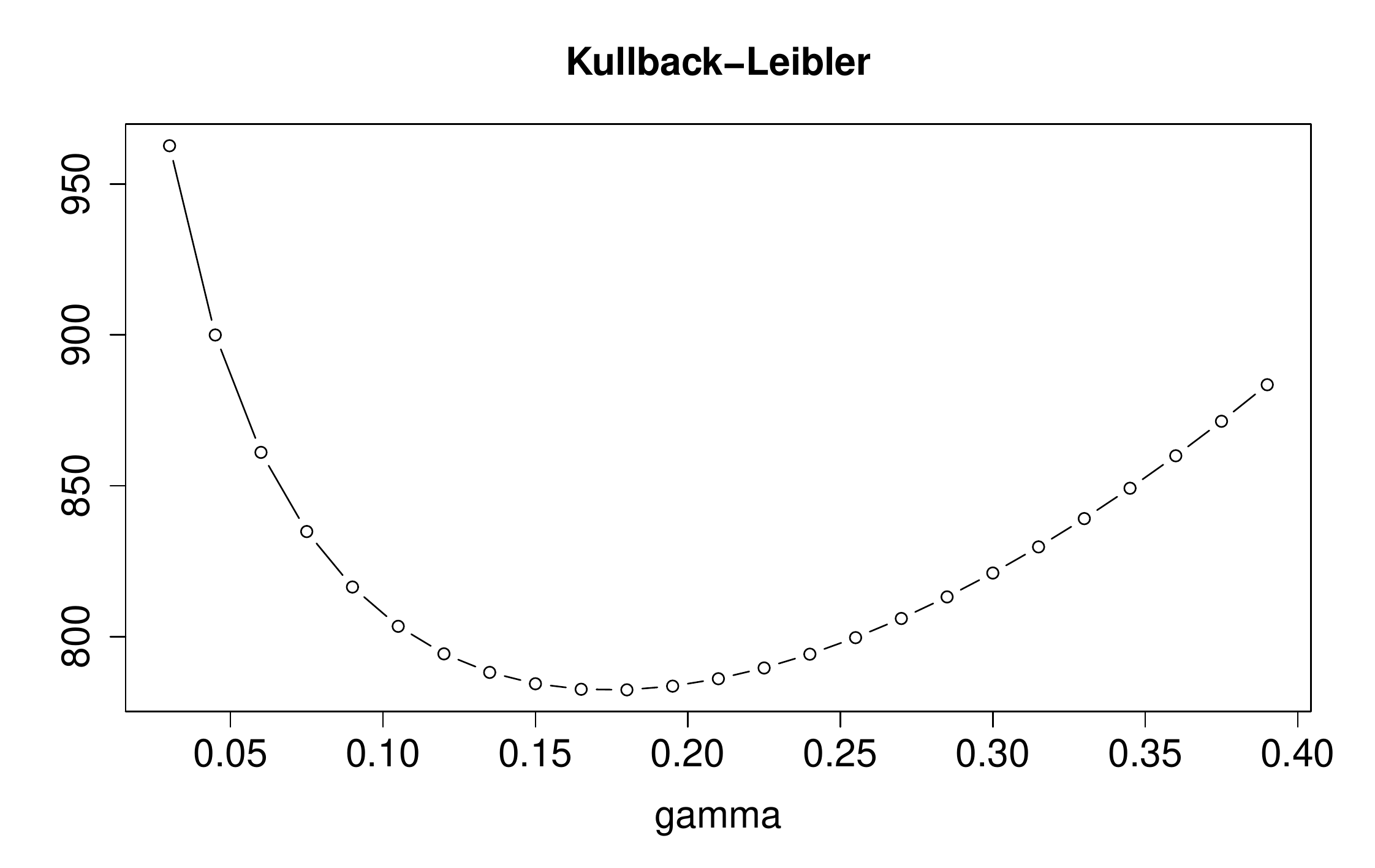}
\caption{Loss functions for the estimation of the scaling factor $\gamma_{10}$ for one simulated data set}.
\label{fig:NV2}                                                  
\end{figure}

For illustration we consider the estimates obtained for the the normal-ogive model, in which $F(.)$ is the normal distribution function. The item parameters for  $I=6$ items were $0, -1.5, -1,   0.5, 1.2, 1.5$, $P=100$ (first row) and $P=300$ (second row). Person parameters were drawn from a standardized normal distribution. Figure \ref{fig:NV1} shows the box plots of estimates (200 repetitions) and the density estimate of item parameter 5.
It is seen that the estimates approximate the true values rather well. It demonstrates that item parameters can be estimated separately also for the normal-ogive model, for which no sufficient statistics exist.
Figure \ref{fig:NV2} shows the  quadratic and the Kullback-Leibler loss functions used to select the scale parameter for the last of the simulated data sets. It is seen that the both loss functions select very similar scale parameters. Although the values of the functions differ the minima for both functions are close to 0.18. In the simulations  the Kullback-Leibler loss has been used.

\subsection{Separation of Parameters in the Binary Rasch Model}                                                                            

Let us again consider the binary Rasch model. As already mentioned in Section \ref{sec:invR} it
 is special among monotone homogeneity models since it allows for estimation of item parameters by using sufficient statistics for person parameters.
The existence of sufficient statistics can be used to maximize the conditional log-likelihood yielding the \textit{conditional estimates}. It also allows to construct further estimates of item parameters that do not depend on the selected persons. 

An alternative estimation method that is strongly linked to the separability seen in equation (\ref{eq:Raschsit}) is conditional likelihood estimation for pairs of items. Under the condition $Y_{pi_1}+Y_{pi_2}=1$ solving the conditional likelihood for observations on items $i_1,i_2$ only yields the estimator 
\begin{align*}
\frac{e^{-\hat\delta_{i_1}}}{e^{-\hat\delta_{i_2}}} = \frac{N_{i_1}}{N_{i_2}},
\end{align*}
where $N_i$  denotes the  number of persons that solved item $i$ and exactly one of the two items $i_1,i_2$. It can be seen as an empirical version of equ. (\ref{eq:Raschsit}) and is referred to as \textit{pairwise conditional estimation}. Item parameters are estimated without reference to the persons involved. A corresponding property  holds for the comparison of persons but is not explicitly given. For more on pairwise estimates see \citet{zwinderman1995pairwise}, \citet{von2016rasch}.  

\begin{figure}[H]
\centering
\includegraphics[width=6.5 cm]{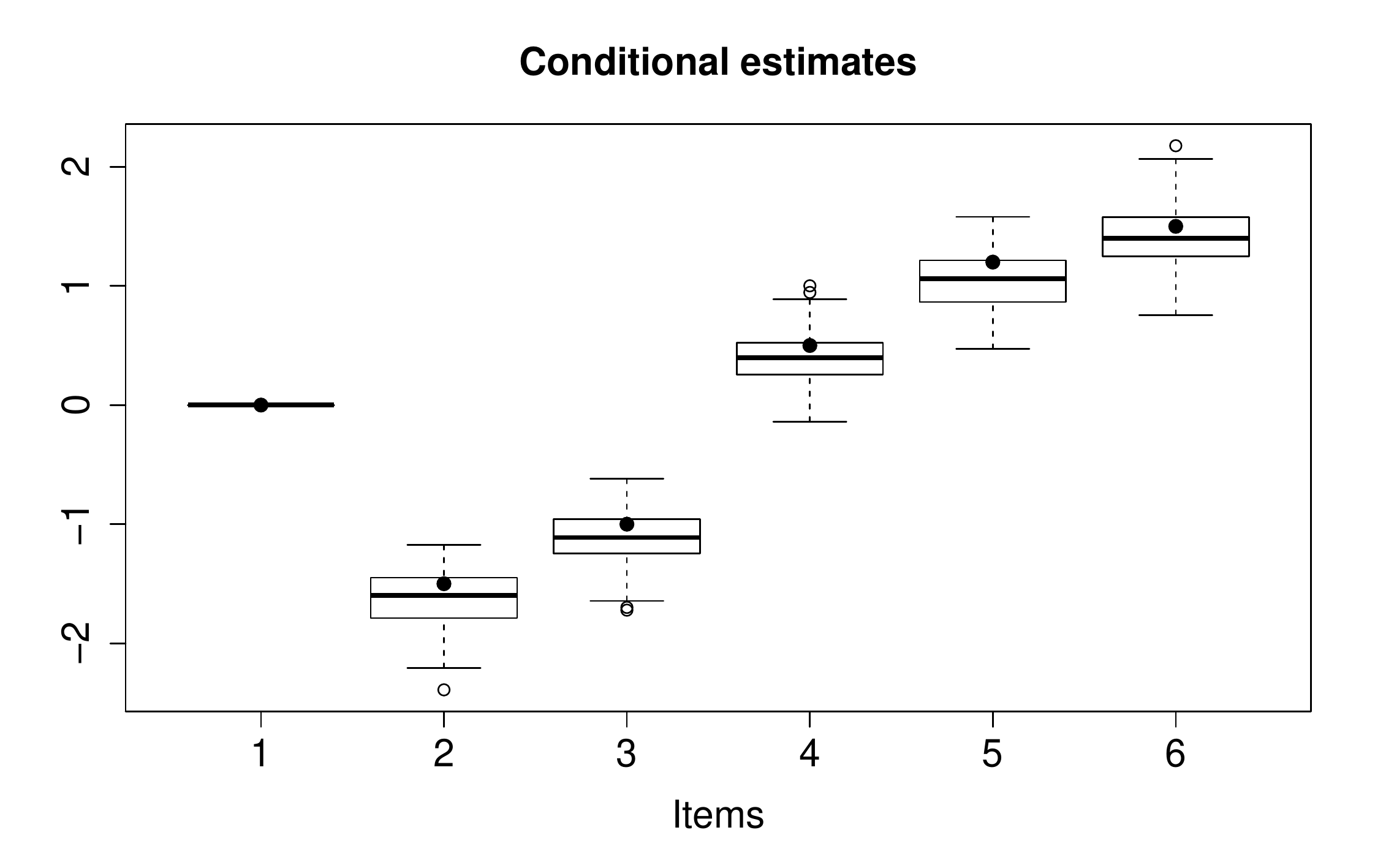}
\includegraphics[width=6.5 cm]{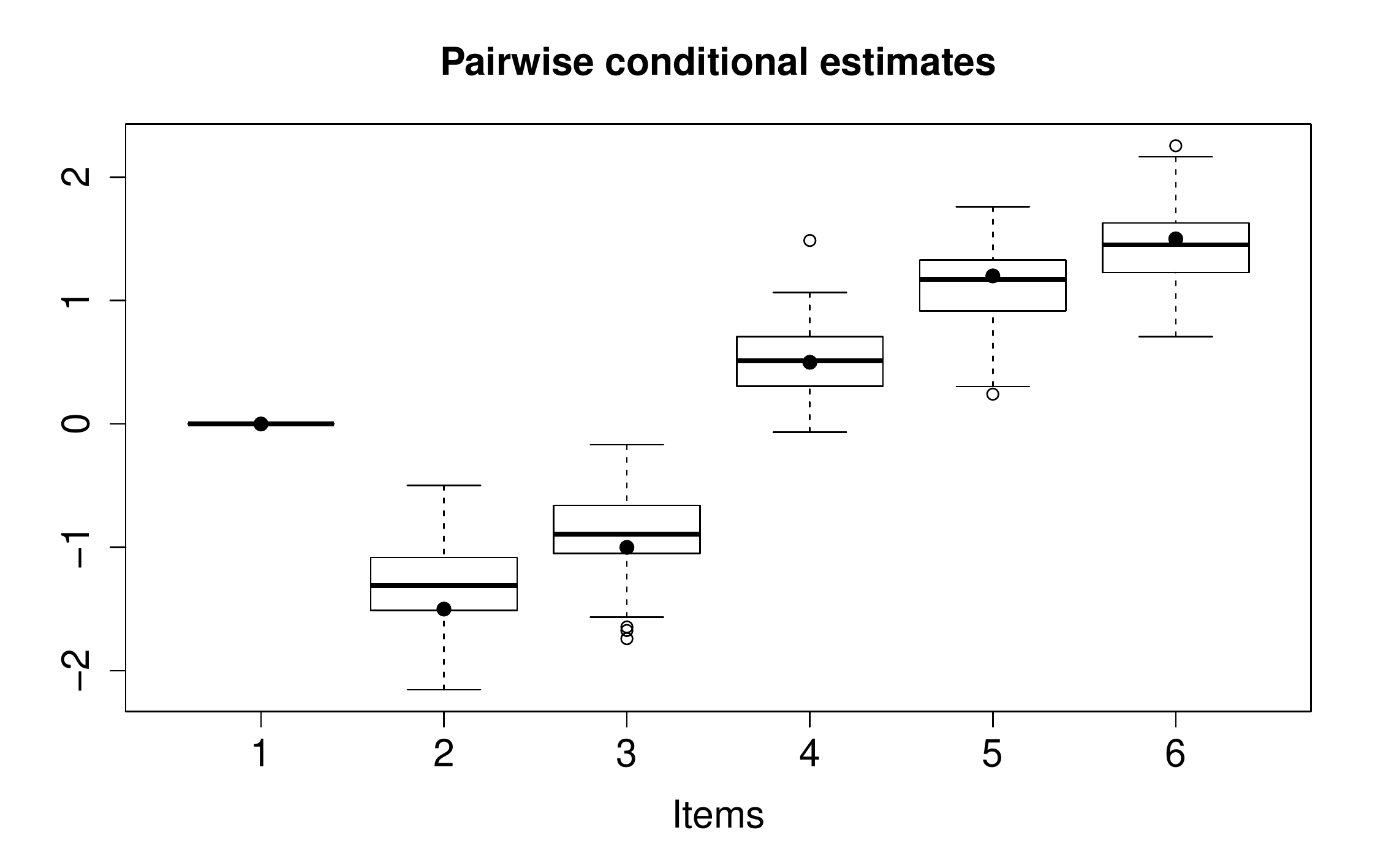}
\includegraphics[width=6.5 cm]{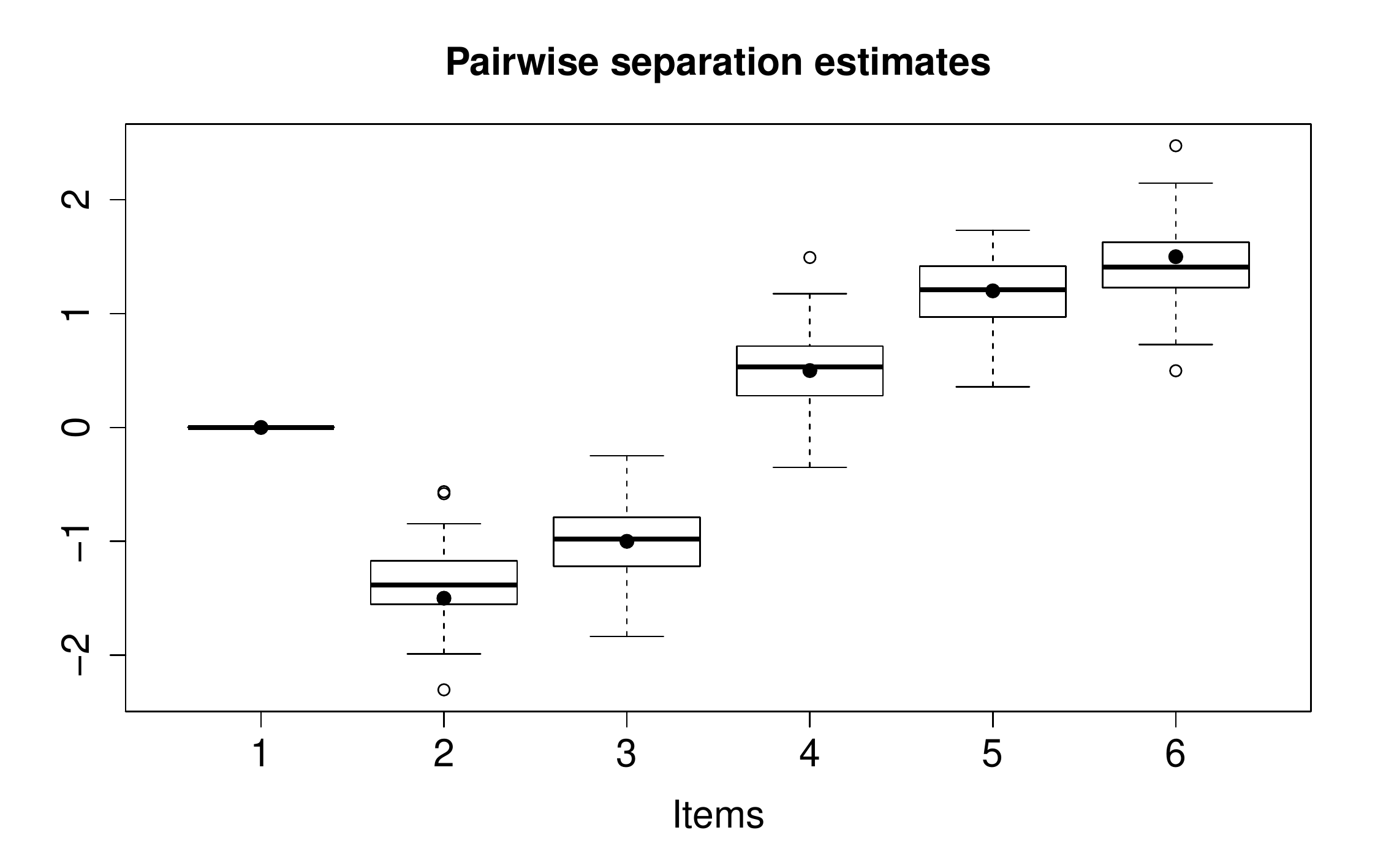}
\caption{Box plots of parameter estimates for conditional, pairwise conditional and separation estimators in the Rasch model with $P=100$}.
\label{fig:nRaschvar100}
\end{figure}

Both methods conditional estimation and pairwise conditional estimation exploit the existence of sufficient statistics given  the total scores, in the latter case under the condition that only two items are considered. In Figure \ref{fig:nRaschvar100} the estimators are compared to the pairwise separation estimator  with the same item parameters as in Figure \ref{fig:NV1} and the assumption that the Rasch model holds.  It is seen that the conditional estimates have smaller variance than the other two estimators. The pairwise conditional estimator and the pairwise separation estimator show almost the same performance. This is also supported by computing the average absolute deviations of estimates from the true values, which was 0.169 for the conditional estimator, 0.215 for the pairwise separation estimator, and 0.217 for the pairwise conditional estimator. Similar pictures are obtained if the number of persons is larger (see appendix).

 
\subsection{Non-symmetric Response Models}

The strength of the pairwise separation estimator is that it provides estimates also when the response function is not the logistic function. That includes response functions that are  not symmetric. 
Non-symmetric skewed response functions are not so often used since item parameters are not so easily obtained as in the Rasch model although there is no compelling reason why item response functions should be symmetric. Advantages of skewed response functions have been outlined by 
\citet{samejima2000logistic}, \citet{bazan2006skew} and, more recently, by \citet{bolt2022item}. 

\begin{figure}[h!]
\centering
\includegraphics[width=6.5 cm]{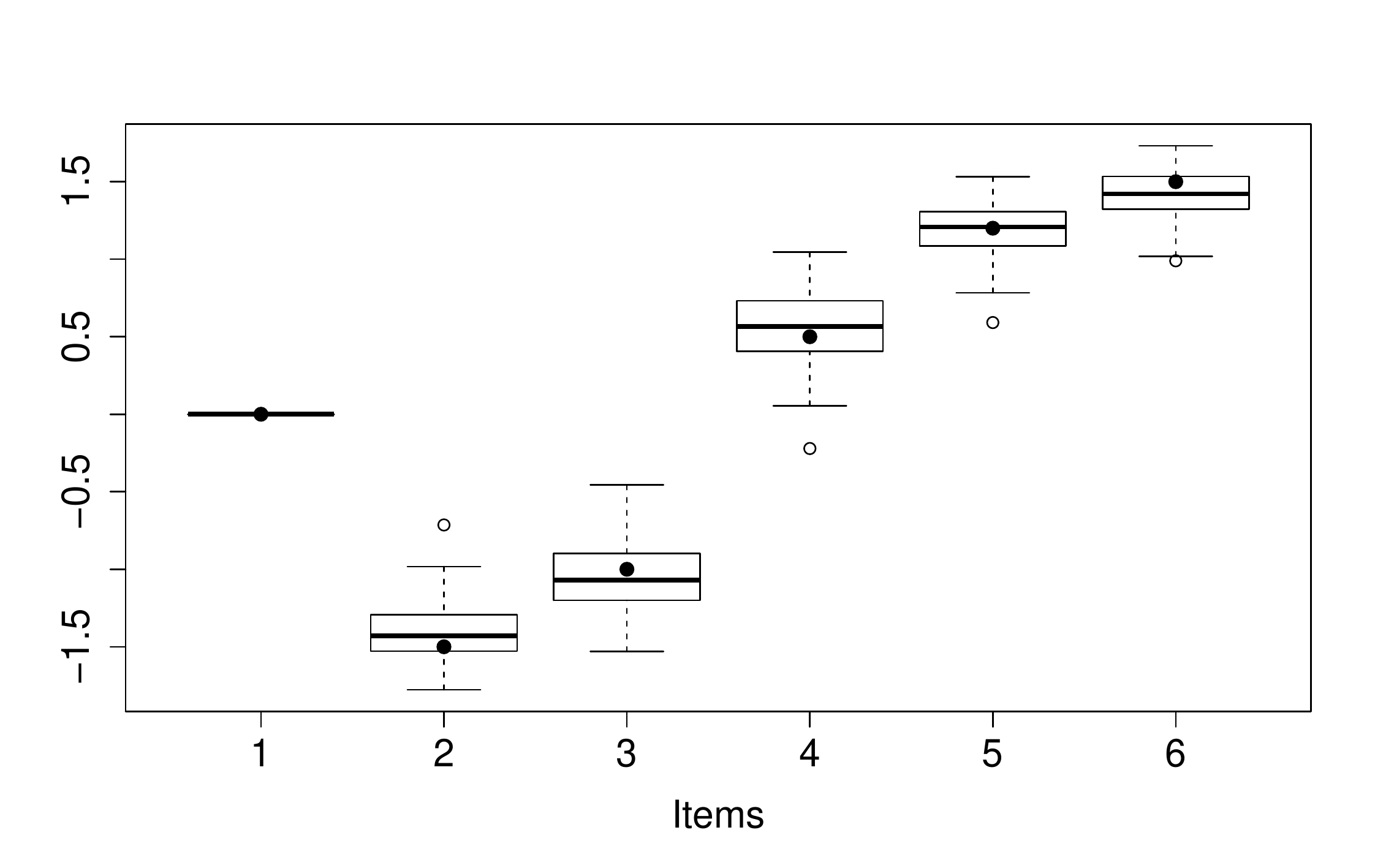}
\includegraphics[width=6.5 cm]{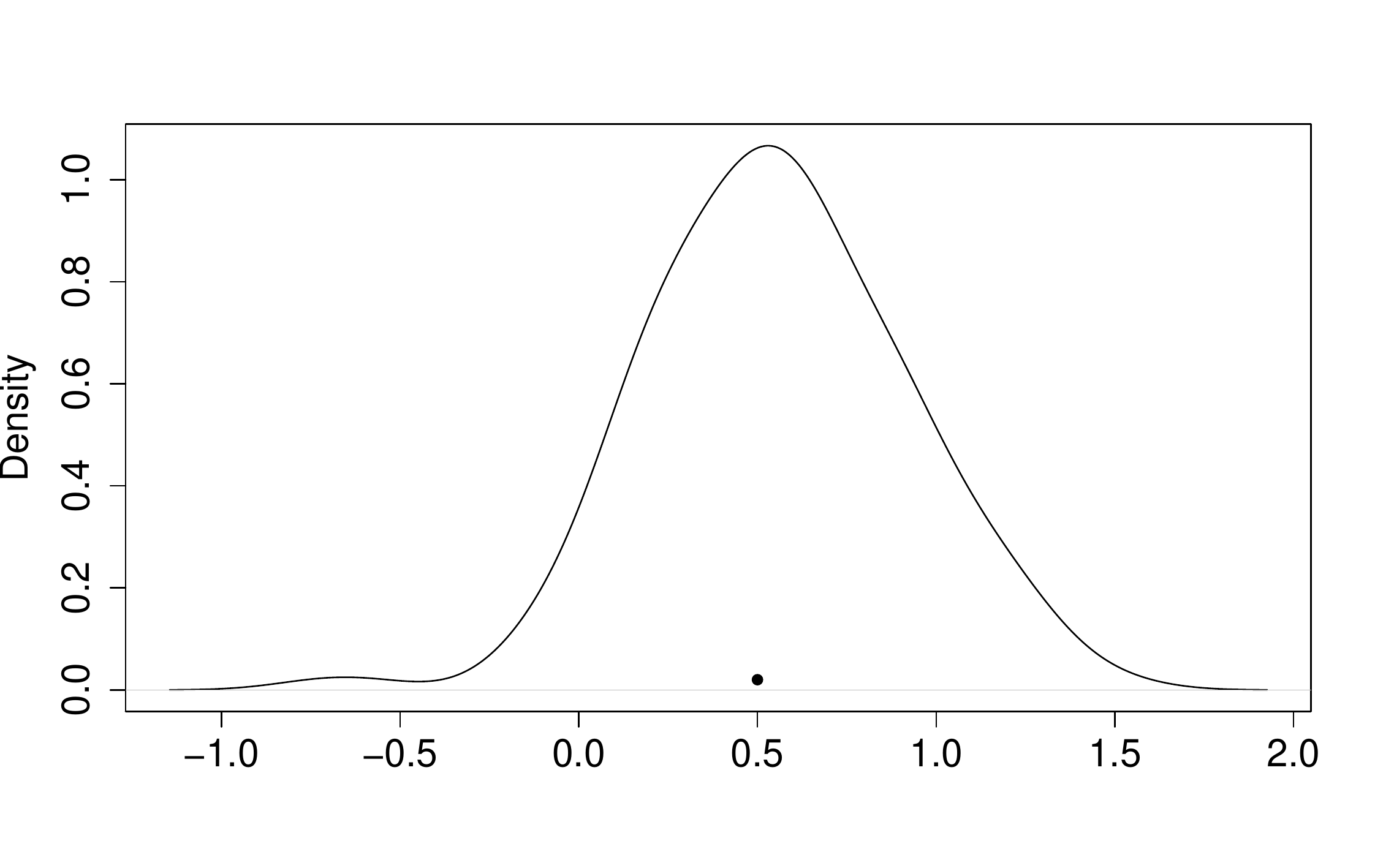}
\caption{Left: Box plots for estimates for six items with dots indicating the true values assuming a maximum value model; right: density of estimates for item 4 $P=100$}.
\label{fig:Gumb1}
\end{figure}

\begin{figure}[H]
\centering
\includegraphics[width=6.5 cm]{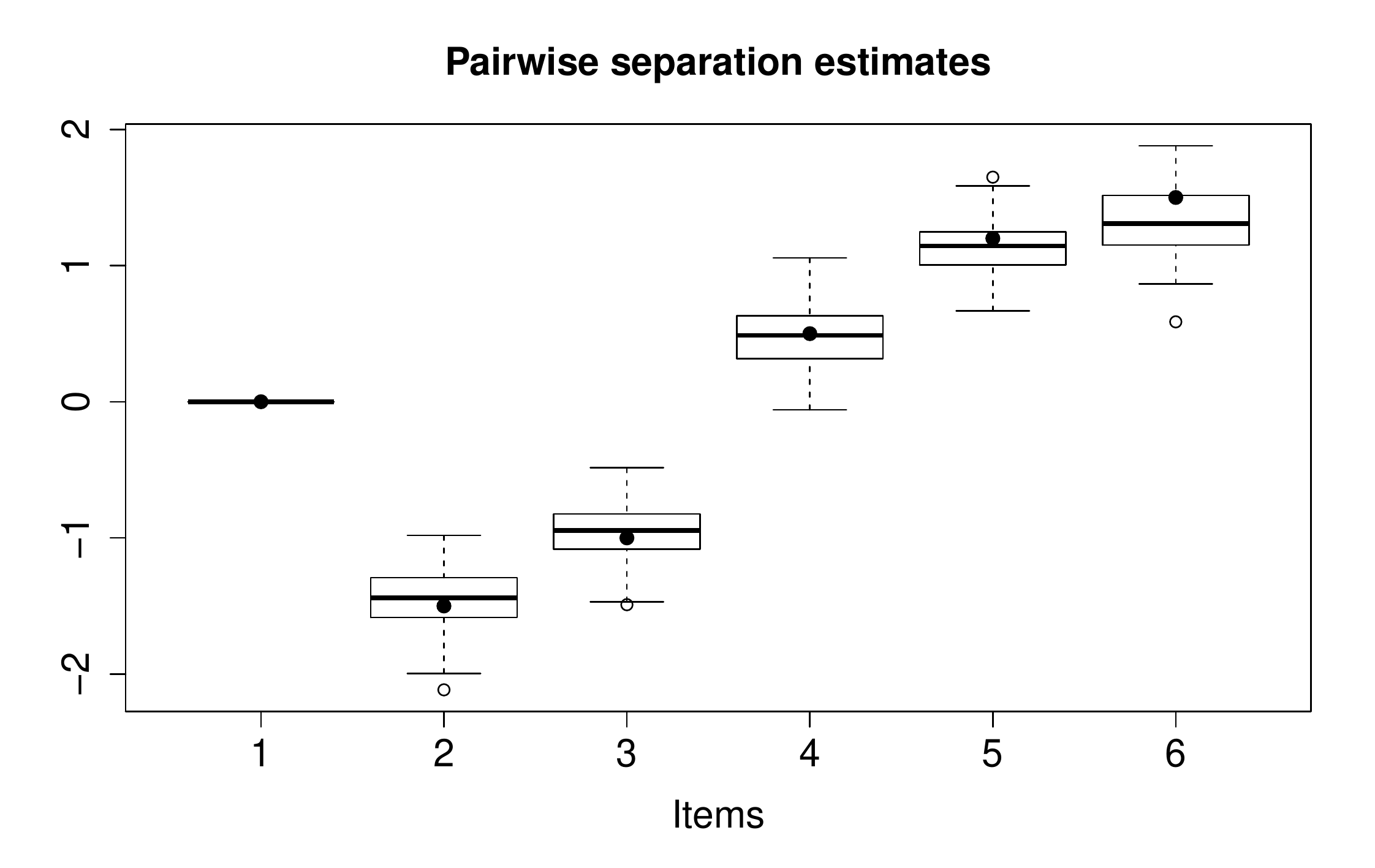}
\includegraphics[width=6.5 cm]{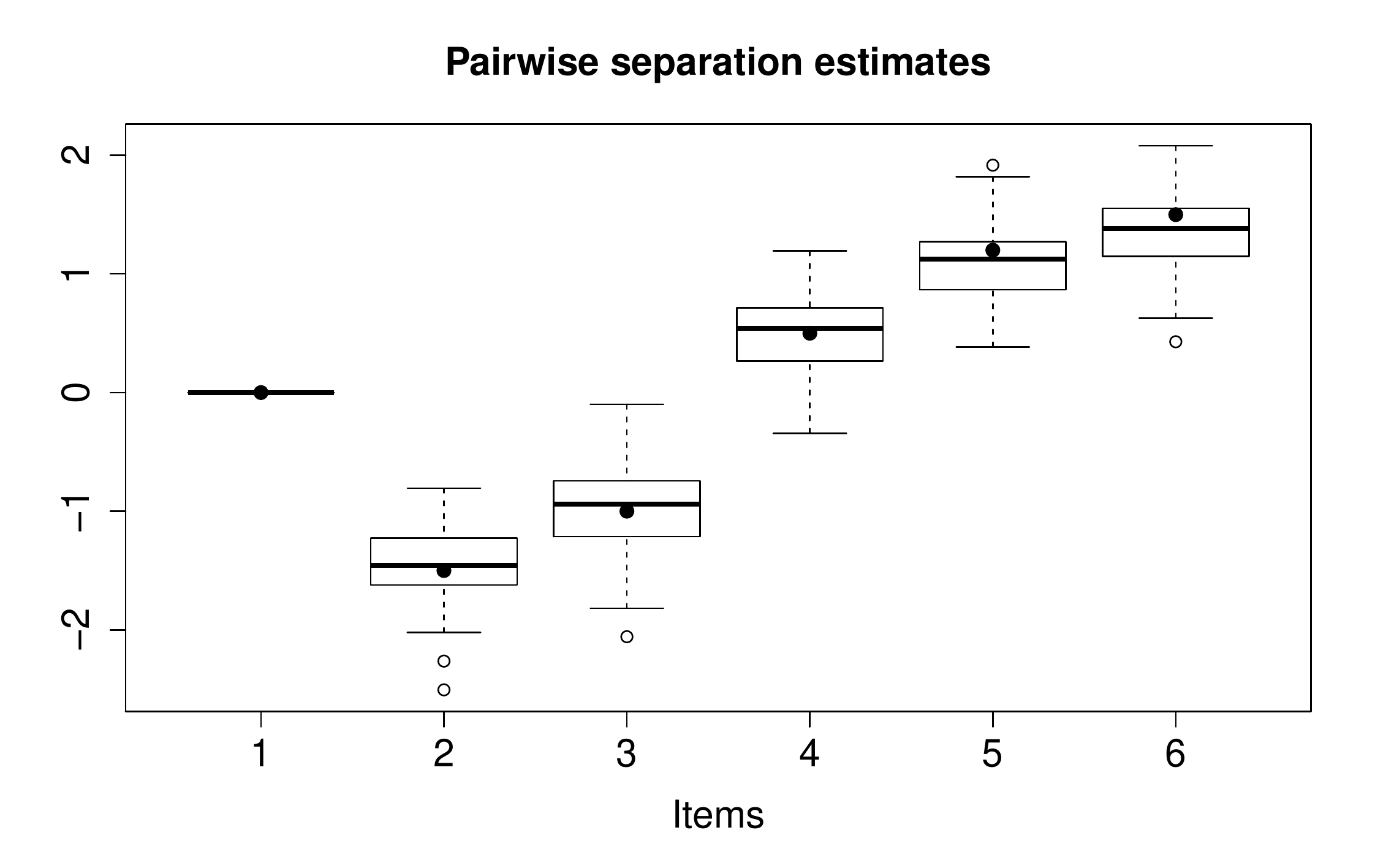}
\caption{Box plots for estimates for six items with dots indicating the true values assuming a minimum value model $P=100$ (left), P=$50$ (right)}.
\label{fig:Gomp1}
\end{figure}

As examples of skewed response functions we consider the maximum value distribution function (Gumbel distribution function) $F(\eta)=\exp(-\exp(-\eta))$ and the minimum value distribution (Gompertz distribution function) $F(\eta)=1-\exp(-\exp(x))$. Figures  \ref{fig:Gumb1} and 
\ref{fig:Gomp1} show the boxplots of estimates when using the pairwise separation estimator (parameters are the same as in Figure \ref{fig:NV1}). Figure   \ref{fig:Gumb1} shows the estimates for the maximum value model ($P=100$) and the density of estimates of item 4 (last simulation). 
Figure \ref{fig:Gomp1} shows the estimates for the minimum value model with $P=100$ on the left and  P=$50$ on the right hand side.
It is seen that the estimator approximates the true parameters rather well.

\subsection{Further Issues}\label{sec:further}

\subsubsection*{Independence of Estimates}
In Rasch models item parameters are often considered to not depend on persons when maximizing the conditional likelihood given person sums
since the conditional likelihood does not contain person parameters. However, this is a misinterpretation. Although the formulae do not contain the person parameters it does not mean that the choice of persons has no impact on the estimates, in particular on their accuracy. It matters, for example, if persons are drawn from a population covering a wide range of abilities or from a population that contains only persons with high (or low) abilities (to be demonstrated in the following). 

Also the separation estimator is constructed by using equations that do not contain the person parameters. Nevertheless, at some point person parameter estimates are used, however only to obtain the scaling parameter since only proportions $\hat\delta_{i}/\hat\delta_{j}$ are scale free. To obtain the scaling parameter $\gamma_{10}$ it is not necessary to use all persons, almost the same estimates are found if only a subset of persons is used to find the scaling parameter. But, as in conditional estimators the selection of the sub population has an impact on the \textit{accuracy} of estimates.

Table \ref{tab:depend} illustrates the effect of the choice of the person population. Person parameters have been  drawn from several distributions,   the standard normal distribution ($\theta_p\sim \NV(0, 1)$), which covers a wide range of person parameters, the $\chi^2$-distribution with one degree of freedom ($\theta_p=\tilde\theta_p^2$ with $\tilde\theta_p\sim \NV(0, 1)$), which means that all person parameters are positive,  the non-central  $\chi^2$-distribution with non-centrality parameter 1 ($\theta_p=\tilde\theta_p^2$ with $\tilde\theta_p\sim \NV(1, 1)$), and  the non-central  $\chi^2$-distribution with non-centrality parameter 1.5 ($\theta_p=\tilde\theta_p^2$ with $\tilde\theta_p\sim \NV(1.5, 1)$). In particular the latter scenarios  mean that only rather large person parameters are in the sample. The item parameters are the same as  in Figure \ref{fig:NV1}. 
As measure of accuracy we use the  absolute deviation $|\delta_i-\hat\delta_i|$ averaged across all items.
Results are given for several sample sizes. It is seen that all estimates perform poorer when the range of person parameters becomes smaller. This includes the   estimators that use the conditional likelihood approach. In particular the pairwise conditional estimator suffers strongly if the available persons are from a smaller range. The separation estimator is relatively stable and performs well also in these cases although for Rasch models as simulated here the conditional estimator shows the best performance. The results demonstrate that also conditional estimators of item parameters can not be considered as being independent of the choice of persons although person parameters are not involved in the estimation equations.

\begin{table}[H]
 \caption{Deviations of estimates of item parameters from true values for several estimation methods} \label{tab:depend}
\centering
\begin{tabularsmall}{lllllllcccccccccc}
  \toprule
  &&pairwise   & conditional & pairwise \\ 
  &&separation   &  & conditional\\
  \midrule
P=80 &standard normal   &0.253    &0.186  &0.289\\
&chi-squared &0.254    &0.184  &0.288\\
&nonc chi-squared, $N(1,1)^2$  &0.288   &0.220  &0.371\\
&nonc chi-squared, $N(1.5,1)^2$   &0.317    &0.262  &0.438\\
\midrule
P=100&standard normal   & 0.215 &0.169  & 0.217              \\
&chi-squared  & 0.210    &0.181  &0.236 \\
&nonc chi-squared, $N(1,1)^2$ &0.256    &0.204  &0.341 \\
&nonc chi-squared, $N(1.5,1)^2$  &0.287  &0.231 &0.378 \\
\midrule                                                              
P=200&standard normal,   &0.167  &0.137 &0.179 \\
&chi-squared       &0.166    &0.135  &0.178 \\
&nonc chi-squared, $N(1,1)^2$   &0.164    &0.145  &0.203 \\
&nonc chi-squared, $N(1.5,1)^2$ &0.184  &0.158 &0.213\\
\bottomrule
\end{tabularsmall}
\end{table}

\subsubsection*{Smoothing and the Separation Estimator}

The separation estimator can be seen as an estimator based on smoothing techniques. Methods as smooth kernel estimators for densities and distribution functions have a long tradition, see, for example,  \citet{WandJones:95}, \citet{Simonoff:96b}.

Let us first consider the general case of a random variable $Y$. For a fixed observation $y_0$ of the random variable the contribution to the density estimate
is the kernel function $g((y-y_0)/h)$, where $g(.)$ is a fixed density function and $h$ is the window width. The contribution to the estimated distribution function is $G((y-y_0)/h)$, where $G(.)$ is the distribution function corresponding to $g(.)$. Alternatively one can consider   estimates of  the survivor function $S(y)= P(Y \ge y)$, which are used in the representation of the binary response models considered here.
 
An un-smoothed step function estimate of $P(Y \ge y)$ when $y_0$ has been observed is the function $S(y)=1-H(y-y_0)$, where $H(.)$ is the Heaviside function 
 $H(x)=1$ if $x \ge 0$ and $H(x)=0$ otherwise. To obtain a smooth estimator one can replace the Heaviside function by a smooth approximation. Any distribution function that is very steep at zero could be used. The choice corresponds to the choice of the kernel in density estimation. For simplicity we choose the function
$F_{\alpha}(\eta)=1-F(-\alpha \eta)$, where $F(.)$ is the response function of the item response model (centered at zero such that $F(0)=.5$) and $\alpha >0$ is a large constant. Since the function is centered at zero for increasing $\alpha$ the function $F_{\alpha}(.)$ becomes the Heaviside function for all values unequal zero.
Smooth estimates result if 
\begin{align*}
&P(Y \ge y) \text{ is estimated by } F(-\alpha(y-y_0)) \text{ and }\\
&F^{-1}(P(Y \ge y)) \text{ is estimated by } F^{-1}(F(-\alpha(y-y_0)))= -\alpha (y-y_0).
\end{align*}
When applied to the responses $Y_{pi}$ one obtains
\begin{align*}
&F^{-1}(P(Y_{pi_1} \ge y)) - F^{-1}(P(Y_{pi_1} \ge y))=\\ 
&=F^{-1}(F(-\alpha(y-y_{pi_1}))-  F^{-1}(F(-\alpha(y-y_{pi_2})) = \\
&=\alpha(y_{pi_1}-y_{pi_2})= 
\begin{cases}                                           
0 &y_{pi_1}=y_{pi_2} \\   
\alpha &y_{pi_1}=1,y_{pi_2}=0\\
-\alpha &y_{pi_1}=0,y_{pi_2}=1\\
\end{cases}
\end{align*}
This is equivalent to the estimate $\hat\delta_{i_2}-\hat\delta_{i_1}$ for one person with the smoothing parameter $\alpha$ corresponding to $\gamma_{10}$.

\subsubsection*{Standard Errors}\label{sec:stan}

A way to obtain  estimates of standard errors is bootstrapping \citep{efron1994introduction, DavHin:97}. For a given data set consisting of $P\times I$ observations, the  estimation procedure is carried out repeatedly for data that are obtained by drawing persons from the data set with replacement. The variation of the estimates is used to compute realistic standard errors of parameters.  

\begin{figure}[H]
\centering
\includegraphics[width=6.5 cm]{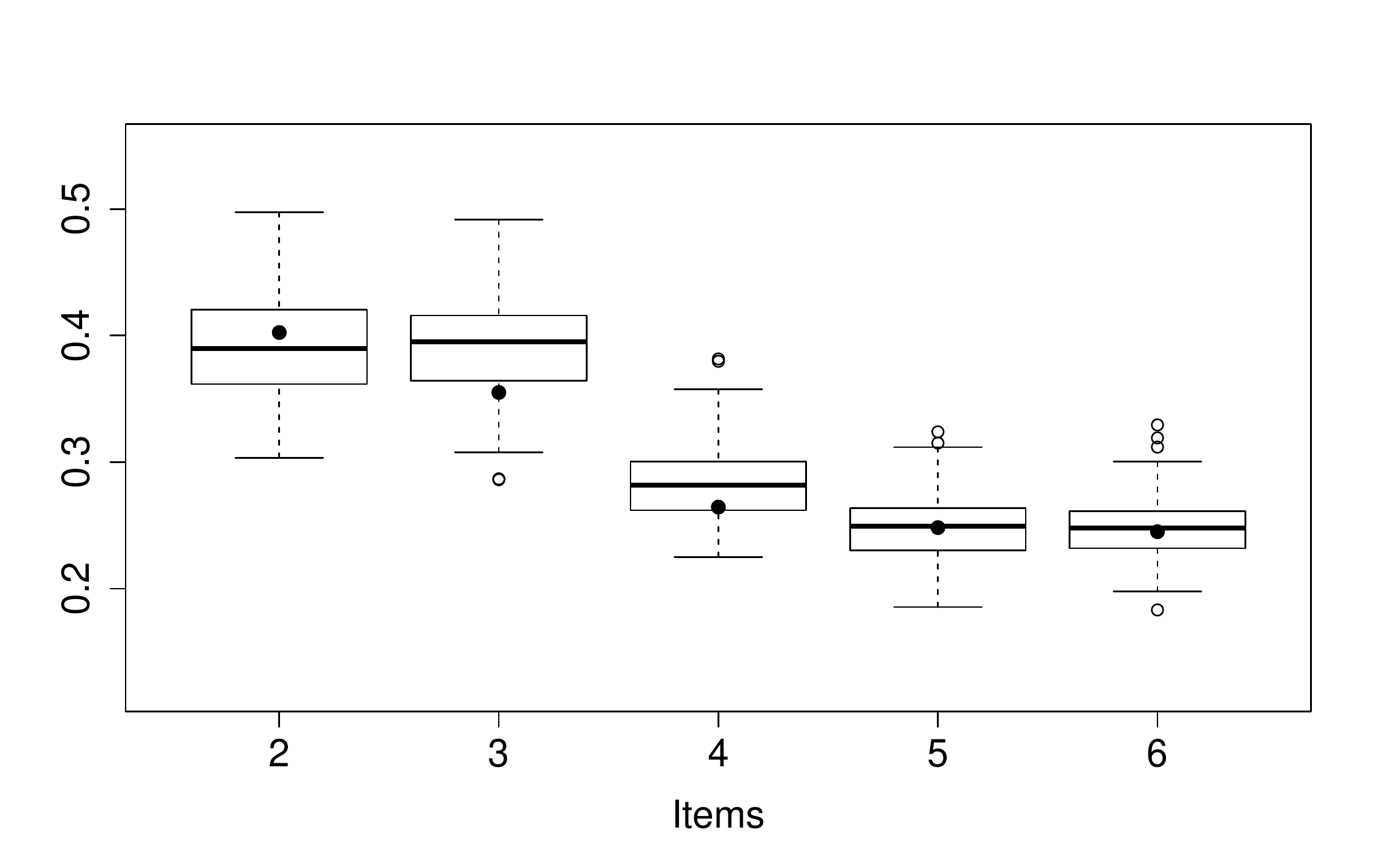}
\caption{Box plots of bootstrap estimators of standard errors, dots denote the ``true''  standard errors (estimated from 300 drawings)}.
\label{fig:boost1}
\end{figure}

For illustration we consider the  approximation of the true standard errors in the scenario given  in Figure \ref{fig:NV1}. Estimates of the true standard errors are computed from   $300$ repetitions. To obtain bootstrap standard errors, we used $50$ repetitions of bootstrap estimates. Each bootstrap estimate is based on $n_{\text{boost}}=200$ drawings with replacement.
Figure \ref{fig:boost1} shows the box plots of bootstrap repetitions. It is seen that bootstrap standard errors approximate the true errors rather well.

\section{Polytomous Models}\label{sec:poly}

Separability of parameters can also be investigated in models with more than two response categories.
Let polytomously scored items have values 
$Y_{pi} \in \{0,1,\dots,k\}$, $p=1,\dots,P$, $i=1,\dots,I$, with  $\{0,1,\dots,k\}$ representing the  coding of  categories.

Several polytomous latent trait models have been considered in the literature, in particular ordinal models as the graded response model \citep{samejima1995acceleration,samejima2016graded} and  the partial credit model \citep{Masters:82,MasWri:84} are widely used. Separability of parameters is especially interesting for the  graded response model since, in contrast to the partial credit model, no sufficient statistics exist that could be used to construct conditional log-likelihood estimators.


\subsection{Invariance in Terms of the Model}

Polytomous models specify the response probabilities 
\begin{equation*} 
\pi_{pir}(\theta_p,\deltab_{i})=P(Y_{pi}=r|\theta_p,\deltab_{i}), \quad r=0,\dots, k,
\end{equation*}  
as functions of person parameters $\theta_p$ and a vector of item parameters $\deltab_{i}$. The latter usually has the form $\deltab_{i}^T=(\deltab_{i1},\dots,\deltab_{ik})$. The total vector of response probabilities for person $p$ and item $i$ is denoted by 
$\pib_{pi}(\theta_p,\deltab_{i})^T=(\pi_{pi0}(\theta_p,\deltab_{i}),\dots,\pi_{pik}(\theta_p,\deltab_{i}))$.
A  widely used polytomous model is the graded response model \citep{samejima1995acceleration,samejima2016graded}, which specifies the \textit{cumulative} probabilities 
\begin{equation*}\label{eq:cum1}
\pi_{pi}(r)=P(Y_{pi} \ge r|\theta_p,\deltab_{i})=F(\theta_p-\delta_{ir}),\quad  r=1,\dots,k,
\end{equation*}
yielding the probabilities $\pi_{pir}(\theta_p,\deltab_{i})=\pi_{pi}(r)-\pi_{pi}(r+1)$.

When considering invariance it is useful to distinguish between person and item parameters since the former are uni-dimensional while the latter are multi-dimensional. 

For a  polytomous latent trait model   \textit{invariance of comparison of person parameters} (\textit{specific objectivity for the comparison of persons}) holds if  a parameterization and at least one transformation function called comparator  $C_{\text{pers}}$ exist such that  for all persons $p_1,p_2$  and all items  

\begin{align*}
C_{\text{pers}}(\pib_{p_1i}(\theta_{p_1},\deltab_{i}),\pib_{p_2i}(\theta_{p_2},\deltab_{i}))) = \theta_{p_1} - \theta_{p_2}. 
\end{align*}

If the cumulative model is the data generating model for each $r>1$ 
\begin{equation}\label{eq:MHcum}
F^{-1}(\pi_{p_1i}(r))-F^{-1}(\pi_{p_2i}(r)))= \theta_{p_1} - \theta_{p_2}  
\end{equation}
holds. Since the cumulative probabilities are simple functions of the vector of probabilities it is obvious that several comparator functions $C_{\text{pers}}$
exist. Equation (\ref{eq:MHcum}) shows that  $k-1$ functions can be used to compare person parameters.

For a  polytomous latent trait model   \textit{invariance of comparison of item parameters} (\textit{specific objectivity for the comparison of items}) holds if  a parameterization and a set of transformation functions (comparators)  $C_{\text{it},i}$ exist such that  for two items $i_1,i_2$ 
the differences of item parameters are \textit{uniquely} determined in the form
\begin{align*}
C_{\text{it},i}(\pib_{pi_1}(\theta_{p},\deltab_{i_1}),\pib_{pi_2}(\theta_{p},\deltab_{i_2})) = \delta_{i_1q}- \delta_{i_2r}, 
\end{align*}
where $i \in \text{Comp}$. In contrast to the comparison of person parameters it is not sufficient to postulate the existence of just one comparator function. Since 
each item has several parameters  more than one comparator function is needed to determine all differences uniquely.
Typically different sets of comparator functions can be found that determine all differences. One could also consider minimal sets of comparator functions but they  would only be of theoretical interest.

If the cumulative model is the data generating model for any $r,q>1$
\begin{align}\label{eq:MHcum1}
F^{-1}(\pi_{pi_1}(r))-F^{-1}(\pi_{pi_2}(q)))= \delta_{i_2q} - \delta_{i_1r}  
\end{align}
holds, which defines the comparator functions.
With the constraint $\delta_{11}=0$ the set of functions 
\begin{align}\label{eq:concrcum1} 
&F^{-1}(\pi_{p1}(r))-F^{-1}(\pi_{p1}(1)))= \delta_{11} - \delta_{1r}, \quad r >1  \nonumber \\
&F^{-1}(\pi_{p1}(1))-F^{-1}(\pi_{pi}(q)))= \delta_{iq} - \delta_{11}, \quad  i >1, q=1,\dots,k,
\end{align}
forms a sufficiently large (and minimal) set of comparator functions such that all differences and therefore item parameters are uniquely determined.

\subsection{Empirical Invariance}

For the investigation of empirical invariance it is helpful to consider split variables, which are  defined by
\begin{equation*}
Y_{pi}{(r)}=\left\{
\begin{array}{ll}
1&Y_{pi}  \ge r \\
0&Y_{pi} < r.
\end{array} 
\right.
\end{equation*}
The split variable are binary variables that  partition the categories  into the subsets $\{0,\dots,r-1\}$ and $\{r,\dots,k\}$. With response categories $\{0,1,\dots,k\}$  one has $k-1$ split variables $Y_{pi}{(1)},\dots,Y_{pi}{(k)}$. Split variables are the empirical analogues to the cumulative probabilities, $Y_{pi}{(r)}$ is an observation of the binary model $P(Y_{pi}{(r)}=1)=P(Y_{pi}  \ge r)$. 
If the cumulative model holds the probability is given by $P(Y_{pi} \ge r)=F(\theta_p-\delta_{ir})$, which yields the binary models
\begin{equation}\label{eq:cumbin}
P(Y_{pi}{(r)}=1)= F(\theta_p-\delta_{ir}), r=1,\dots,k.
\end{equation}
The link between split variables and the cumulative model is even stronger since the cumulative model holds if models (\ref{eq:cumbin}) hold simultaneously for $r=1,\dots,k$, see also \citet{Tu2020JMP}.

For the cumulative model the theoretical invariance of items is formulated in equ. (\ref{eq:MHcum1}). The empirical analogue is obtained by replacing 
the cumulative probabilities by their observational counterparts. However, as for binary models instead of using the observations themselves one uses pseudo observations, which are slightly distorted original observations. Thus, instead of $Y_{pi}{(r)}$ the pseudo observation 
$Y_{pi}{(r)}^{*}= Y_{pi}{(r)}(1-2\gamma)+\gamma$ is used yielding  
\[
F^{-1}(Y_{pi_1}(r)^{*})-F^{-1}(Y_{pi_2}(q)^{*}))= \begin{cases}
0 &Y_{pi_1}(r)= Y_{pi_2}(q)\\   
\gamma_1- \gamma_0&Y_{pi_1}(r)=1, Y_{pi_2}(q)=0 \\
\gamma_0- \gamma_1&Y_{pi_1}(r)=0, Y_{pi_2}(q)=1,\\
\end{cases}
\]
where $\gamma_0=F^{-1}(\gamma), \gamma_1=F^{-1}(1-\gamma)$. It is an estimator of $\delta_{i_2q} - \delta_{i_1r}$. 
From this one can  construct the estimator
\begin{align}
&\hat\delta_{i_2q} - \hat\delta_{i_1r}= \nonumber\\
&= \frac{(\gamma_1- \gamma_0) n(Y_{pi_1}(r)=1, Y_{pi_2}(q)=0)+(\gamma_0- \gamma_1) n(Y_{pi_1}(r)=0, Y_{pi_2}(q)=1)}{n(Y_{pi_1}(r) \ne Y_{pi_2}(q))} \nonumber \\
&= \frac{(\gamma_1- \gamma_0)(n(Y_{pi_1}(r)=1, Y_{pi_2}(q)=0)-n(Y_{pi_1}(r)=0, Y_{pi_2}(q)=1)}{n(Y_{pi_1}(r) \ne Y_{pi_2}(q)),}\label{eq:genpoly}
\end{align}
where $n(Y_{pi_1}(r)=r, Y_{pi_2}(q)=s)$ is the number of observations with $Y_{pi_1}(r)=r, Y_{pi_2}(q)=s$. Using the reduced set of differences given in equ. (\ref{eq:concrcum1}), which means setting $i_1=r=1$ and  $\delta_{11}=0$, yields
\begin{align}\label{eq:single}
&\hat\delta_{iq}= \frac{(\gamma_1- \gamma_0)(n(Y_{p1}(1)=1, Y_{pi}(q)=0)-n(Y_{p1}(1)=0, Y_{pi}(q)=1)}{n(Y_{p1}(1) \ne Y_{pi}(q))}.
\end{align}
The choice of the scaling parameter $\gamma_{10}=\gamma_1- \gamma_0$ is again based on minimization of a loss function. The quadratic loss 
has the form $L_Q(\yb_{pi},\pib_{pi}(\hat\theta_{p},\hat\deltab_{i}))=\sum_r (y_{pir}-\pib_{pir}(\hat\theta_{p},\hat\deltab_{i}))^2$, where
$\yb_{pi}^T=(y_{pi0},\dots,y_{pik})$, $y_{pir}=1$ if $Y_{pi}=r$,  $y_{pir}=0$ otherwise, and  $\pib_{pir}^T=(\pi_{pi0},\dots,\pib_{pik})$,                 
$\pi_{pir}=P(Y_{pi}=r)$. The Kullback-Leibler  loss is given by $L_{KL}(\yb_{pi},\pib_{pi}(\hat\theta_{p},\hat\deltab_{i}))=\sum_r y_{pir}\log(y_{pir}/\pib_{pir}(\hat\theta_{p},\hat\deltab_{i})$.  

The estimator (\ref{eq:single}) does not use the available information efficiently since it uses only separator functions that refer to item 1 as anchor item
(by choosing $i_1=r=1$ in equ. (\ref{eq:genpoly})). More efficient estimates are obtained by using all items as anchor items one at a time. Thus the final  estimator is constructed as an average over these estimates. It can be computed in an easy way by swapping items. For each pairs of items $(1,i)$ one computes the estimator for ``new'' observations $\tilde Y_{pi}$, where $\tilde Y_{p1}= Y_{pi}$, $\tilde Y_{pi}= Y_{p1}$, and $\tilde Y_{pj}= Y_{pj}$ for $j \ne 1,j \ne i$. This yields $I$ sets of estimated item parameters that are averaged.

\begin{figure}[H]
\centering
\includegraphics[width=6.5 cm]{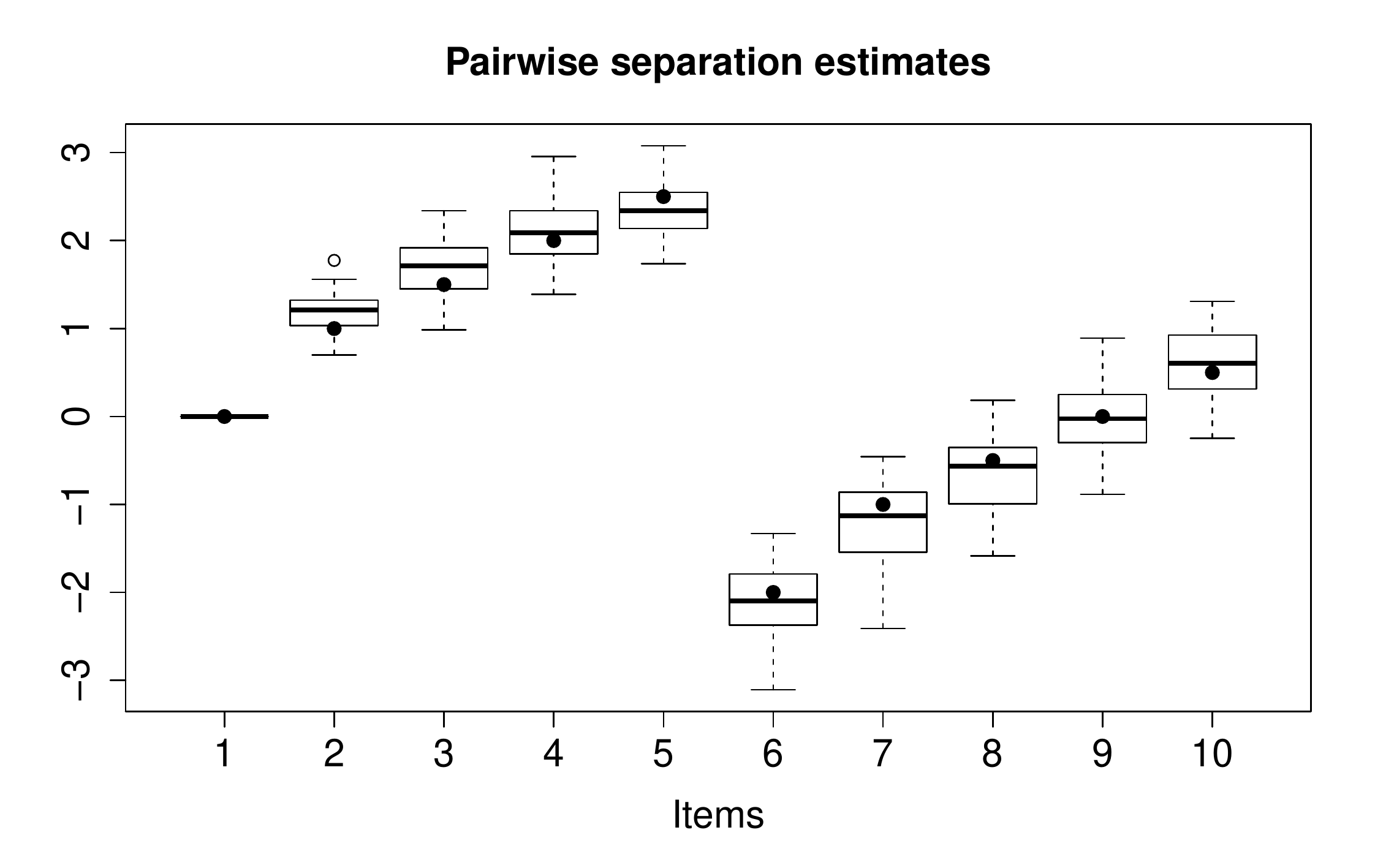}
\includegraphics[width=6.5 cm]{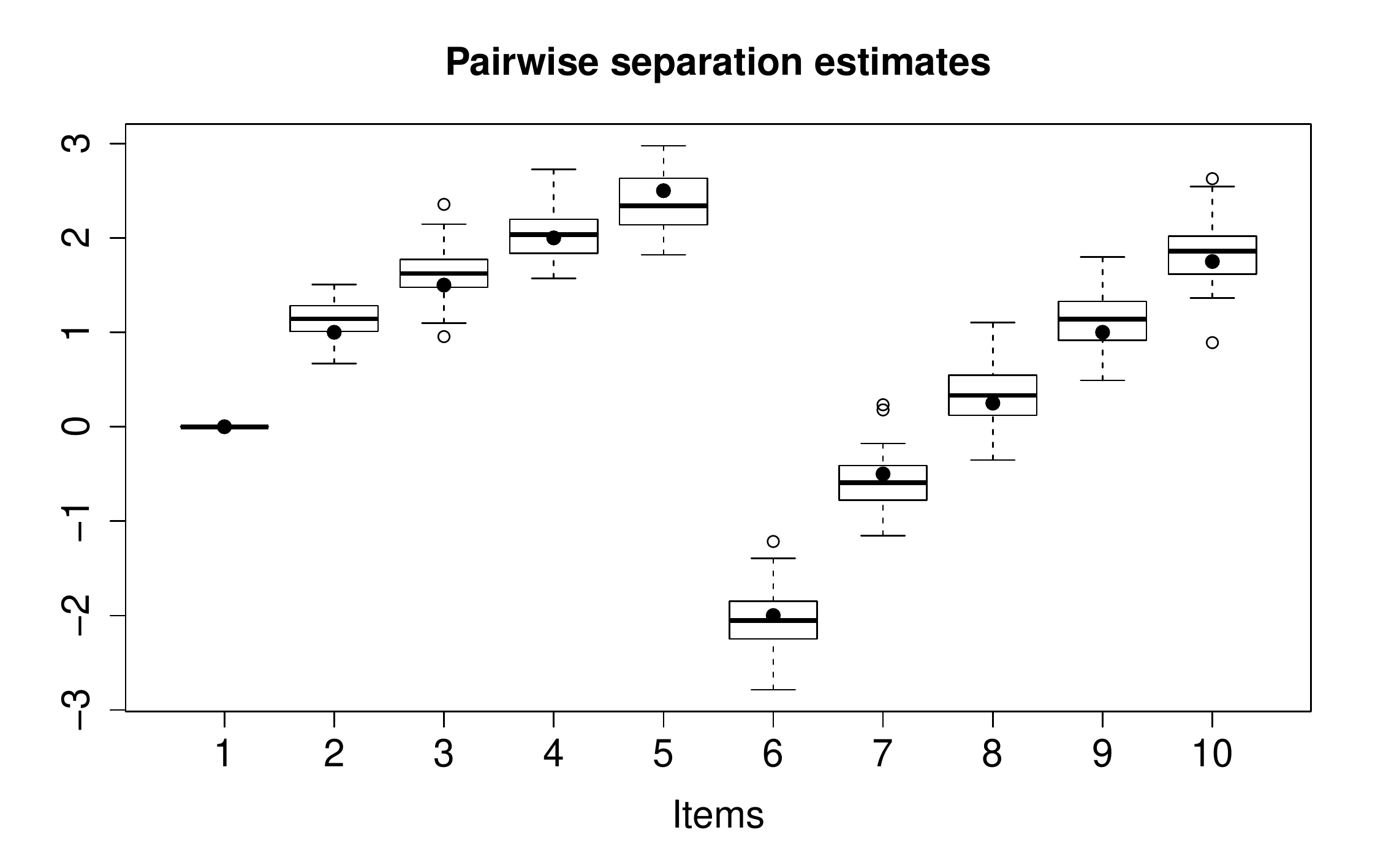}
\includegraphics[width=6.5 cm]{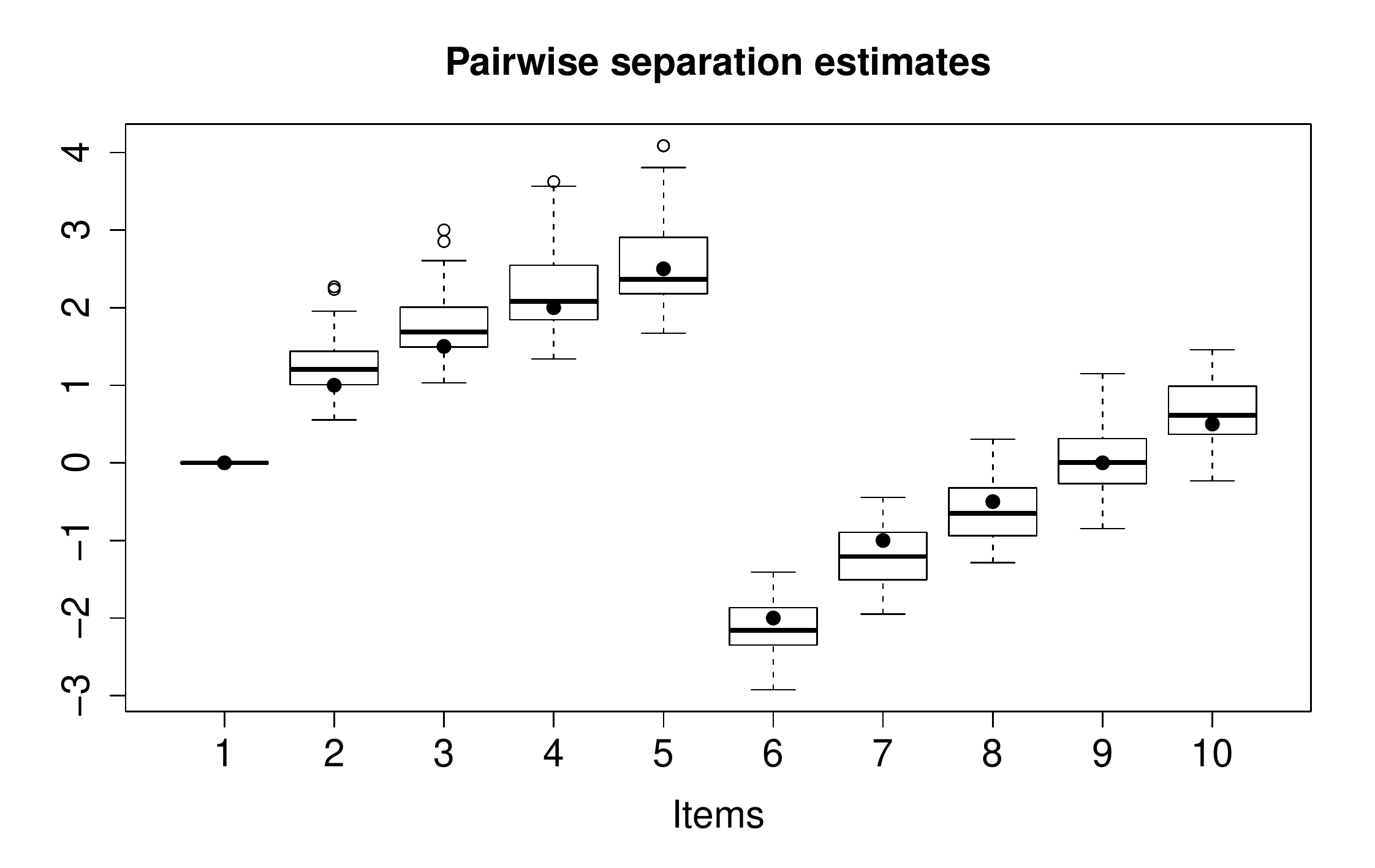}
\includegraphics[width=6.5 cm]{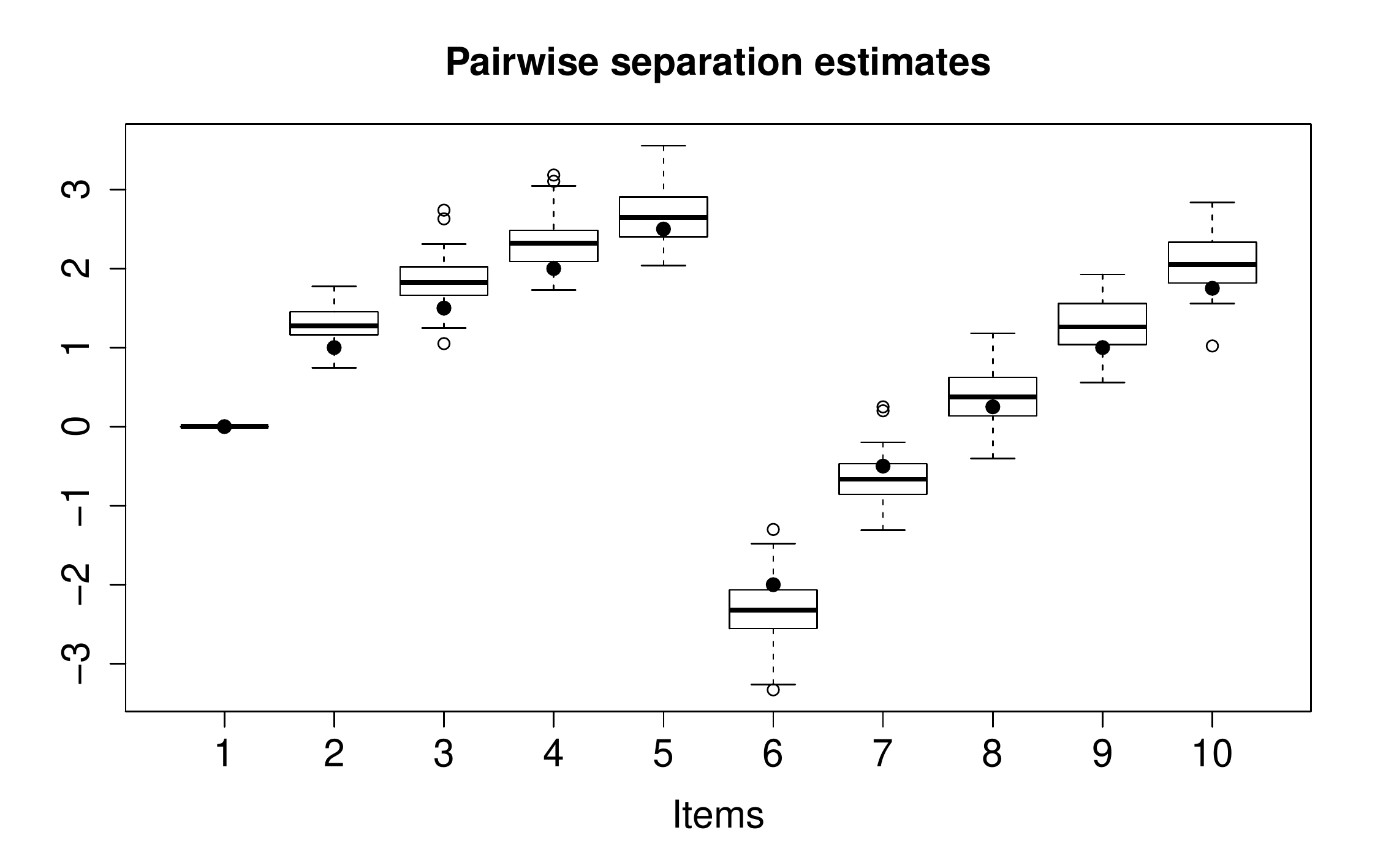}
\caption{Cumulative logit model, $P=100, I=4, k=5$, only first two items, left: $(\delta_{11},\dots,\delta_{15})=((0,1, 1.5, 2, 2.5 ))$, other items are shifted versions, right: second item modified, first row shows the averaged estimator, second row shows the estimator with anchor item 1}.
\label{fig:cum1}
\end{figure}

\begin{figure}[H]
\centering
\includegraphics[width=6.5 cm]{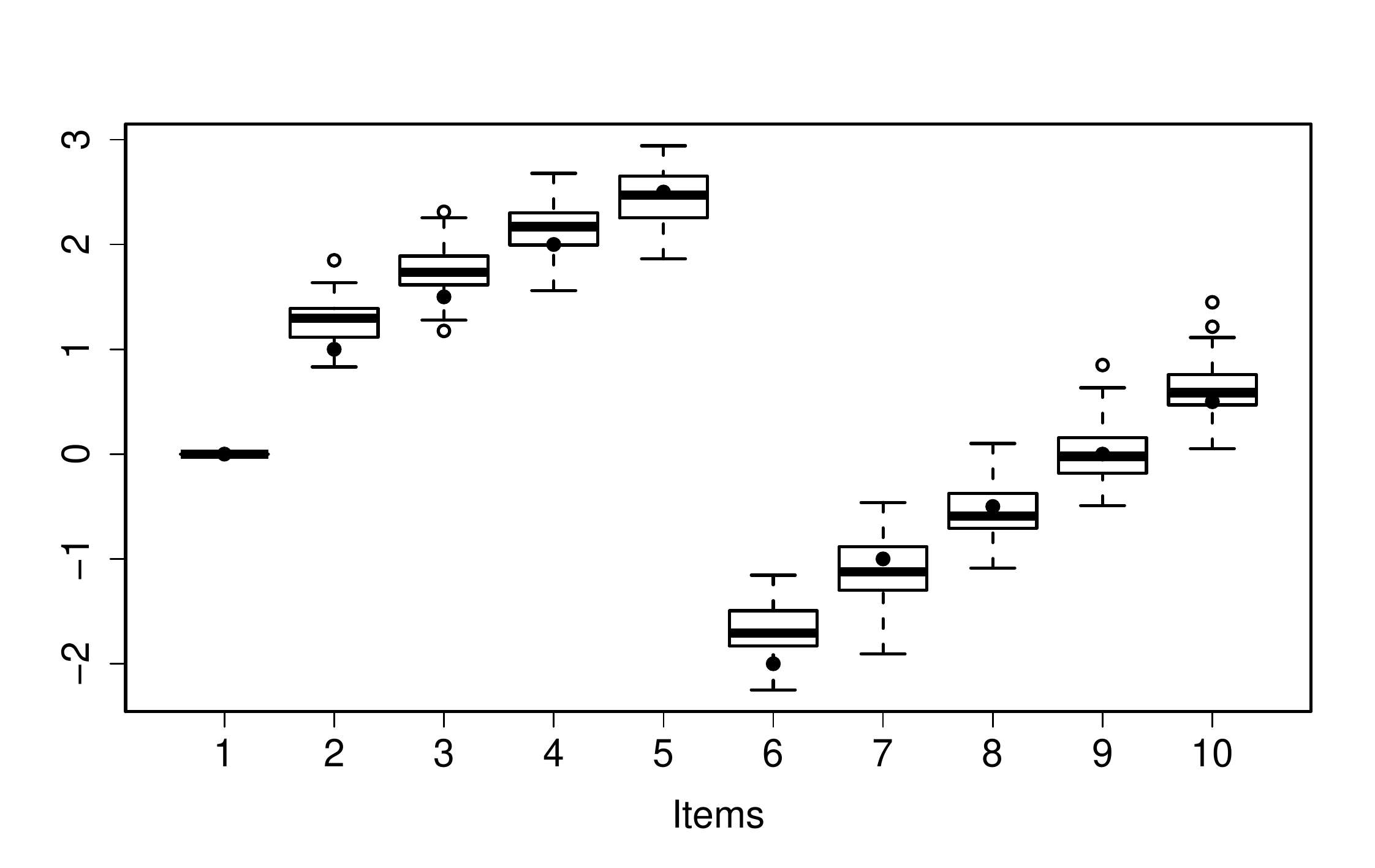}
\includegraphics[width=6.5 cm]{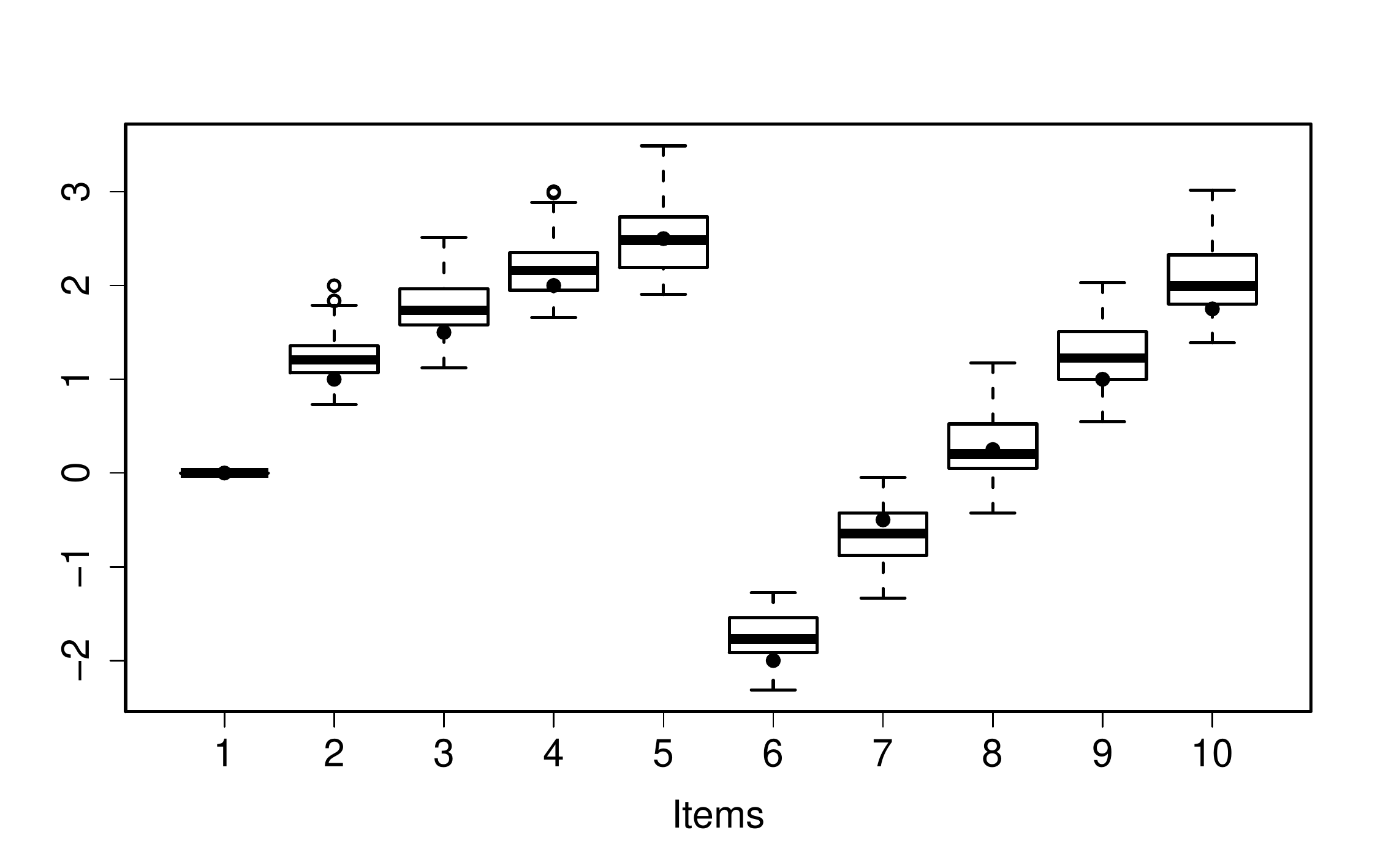}
\caption{Cumulative minimum value model, $P=100, I=4, k=5$, only first two items, left: $(\delta_{11},\dots,\delta_{15})=((0.0 1.0 1.5 2.0 2.5 ))$, other items are shifted versions, right:second item modified}.
\label{fig:cumGomp}
\end{figure}

For illustration Figure \ref{fig:cum1} shows the resulting estimates for the logistic cumulative model with $k=5$, $I=4$, $P=100$ (only first two items shown).
The item parameters are given by $(\delta_{11},\dots,\delta_{15})=((0,1, 1.5, 2, 2.5 ))$. The other items are shifted versions (-2 for item 2, -1.5 for item 3, -3 for item 4). In the right picture item 2 is modified by $(\delta_{21},\dots,\delta_{25})=1.5 (\delta_{11},\dots,\delta_{15})$.
The first row shows the resulting box plots for the averaged estimator, in the second row only item 1 is considered as anchor item.
It is seen that in particular for the averaged estimator the true parameters (dots) are approximated rather well. The corresponding results when fitting the model with the minimum value distribution as response function (averaged estimator) are given in Figure \ref{fig:cumGomp}.

\section{ Concluding Remarks} 
It has been demonstrated that parameters are separable   in a wider class of models, which holds for binary as well as polytomous latent trait models. To this end an  estimator is derived that uses pseudo observations and can be derived as a smoothing estimator. We do not claim that this is the only possible or best estimator. There might be much better ones that estimate parameters without reference to the other group of parameters. 

The basis for success in deriving such an estimator is that separability as a property of the theoretical measurement model holds. If it holds it should be possible to derive a corresponding estimator. The main reason why only Rasch models are considered to allow for separability is that   (conditional) maximum likelihood estimation has been considered the only option to generate estimators that separate parameters. However, there is no general reason why maximum likelihood estimation has to be used. Restriction to maximum likelihood estimation as a principle to obtain empirical separability is an unnecessarily  limited view. 

Separability in the measurement model is based on the existence of comparator functions of the form
$C(\pib_{pi},\pib_{qj})$ with two probability vectors as arguments. More general functions with more arguments could be used to separate parameters. For example, \citet{irtel1995extension} proposed  a function that uses four arguments to compare three subjects in a model with slopes. His function could be seen as a comparator function in the sense used here. However, he considered theoretical comparability only and did not propose an estimator of parameters. 

For the separation estimator software will be made available on Github. Conditional likelihood estimators in binary Rasch models have been obtained by using the eRm package \citep{mair2009extended}.

Thanks to Pascal Jordan and Clemens Draxler for their helpful comments on an earlier draft of the paper.

\bibliography{literatur}

\section{Appendix}

Some more simulation results are given that demonstrate the efficiency of the separation estimator. Figure \ref{fig:Raschvar100} shows the performance of several estimators in a simulation with 18 items. The data generating model is the binary Rasch model, dots denote the true parameters.

\begin{figure}[H]
\centering
\includegraphics[width= 6.5 cm]{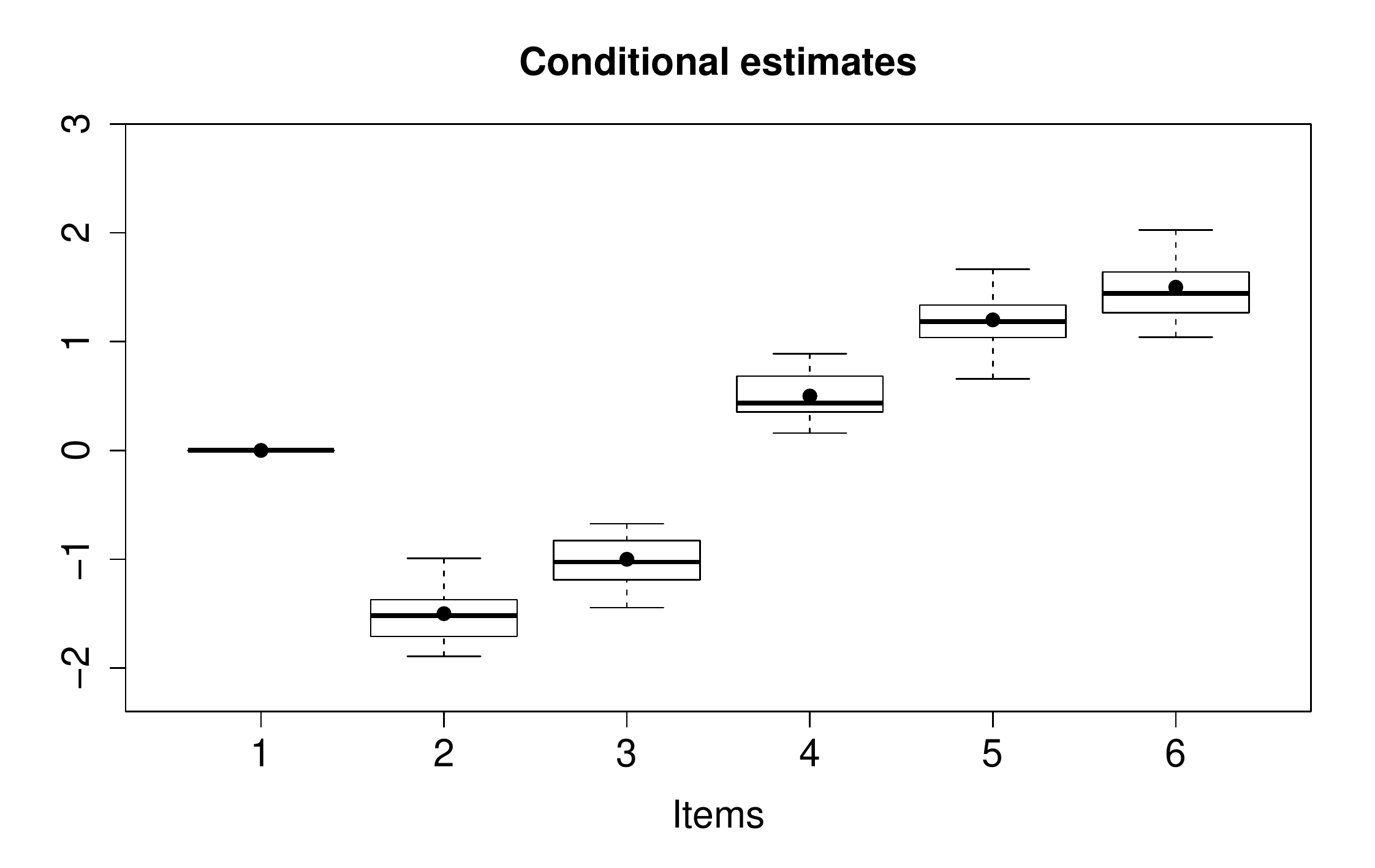}
\includegraphics[width=6.5 cm]{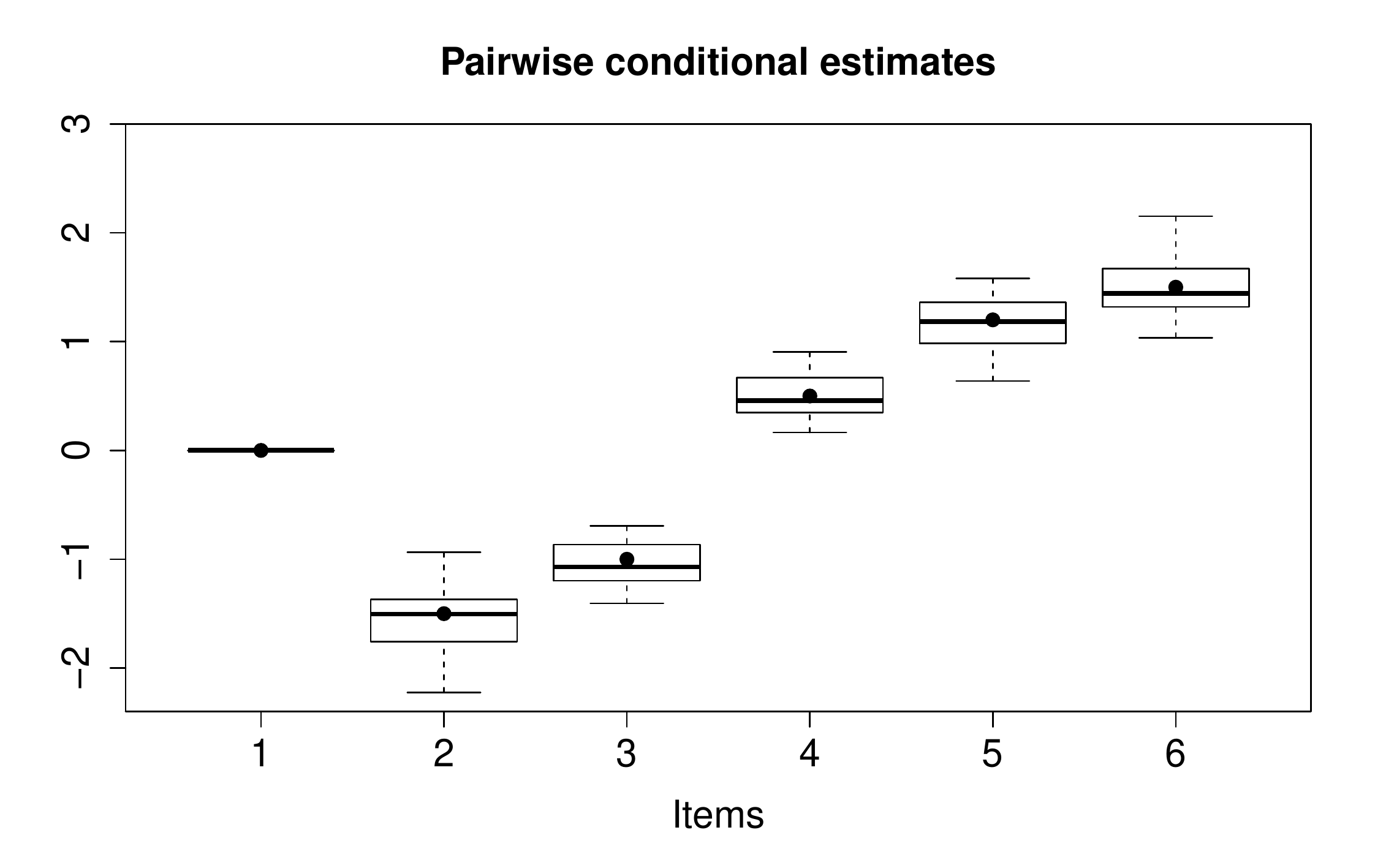}
\includegraphics[width=6.5 cm]{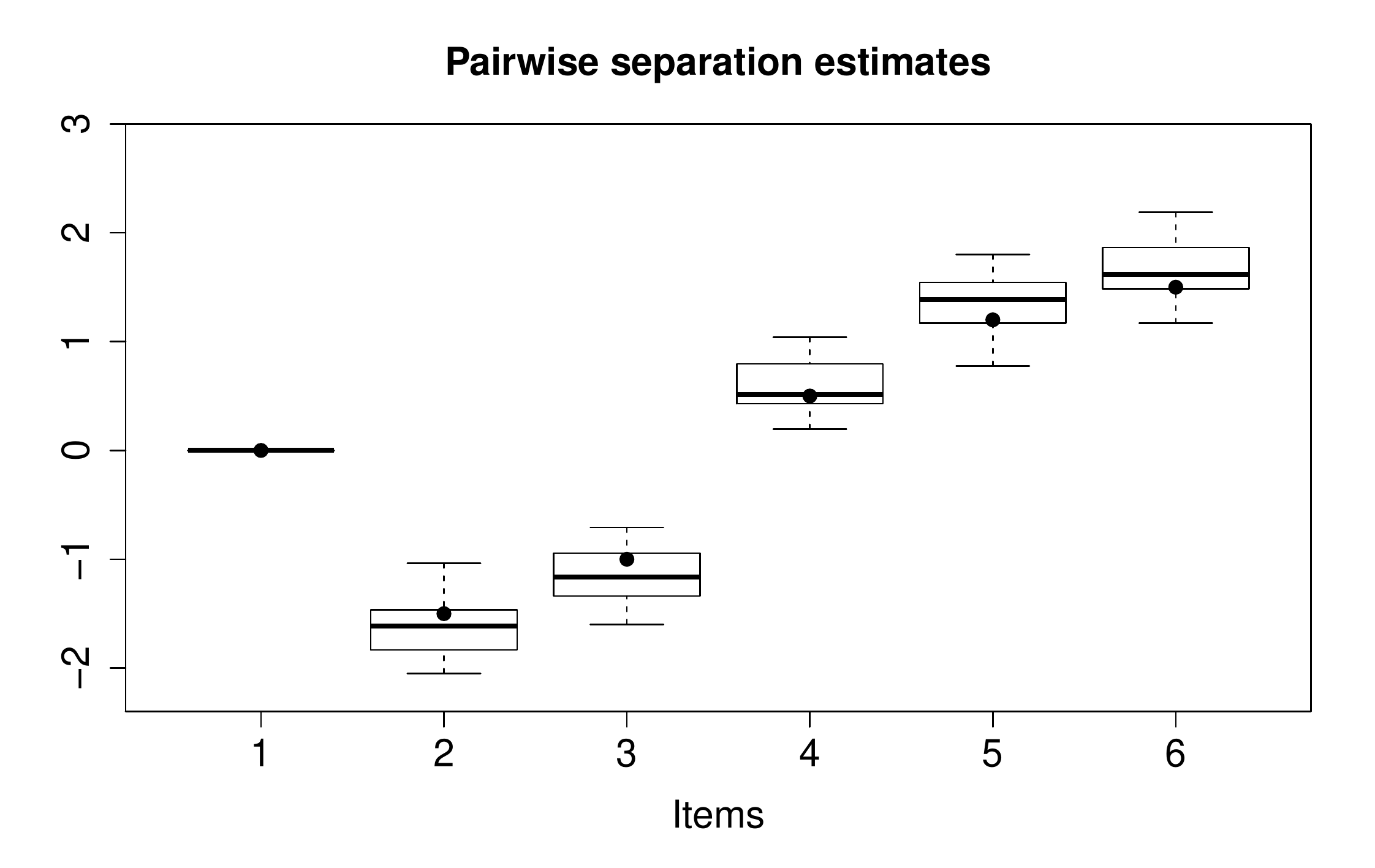}
\caption{Box plots of parameter estimates for conditional, pairwise conditional and separation estimators in the Rasch model with $P=300$}.
\label{fig:Raschvar100n}
\end{figure}

Figure \ref{fig:Raschvar100n} shows the box plots of estimates for the scenario considered in Figure \ref{fig:nRaschvar100} but with larger number of person parameters.

Figure \ref{fig:Raschvar100} shows the box plots of estimates for the binary Rasch model with $I=18$ items, $P=100$ persons drawn from $N(0,1)^2$. 
It is seen that the performance of the  estimators is rather similar.

\begin{figure}[H]
\centering
\includegraphics[width=6.5 cm]{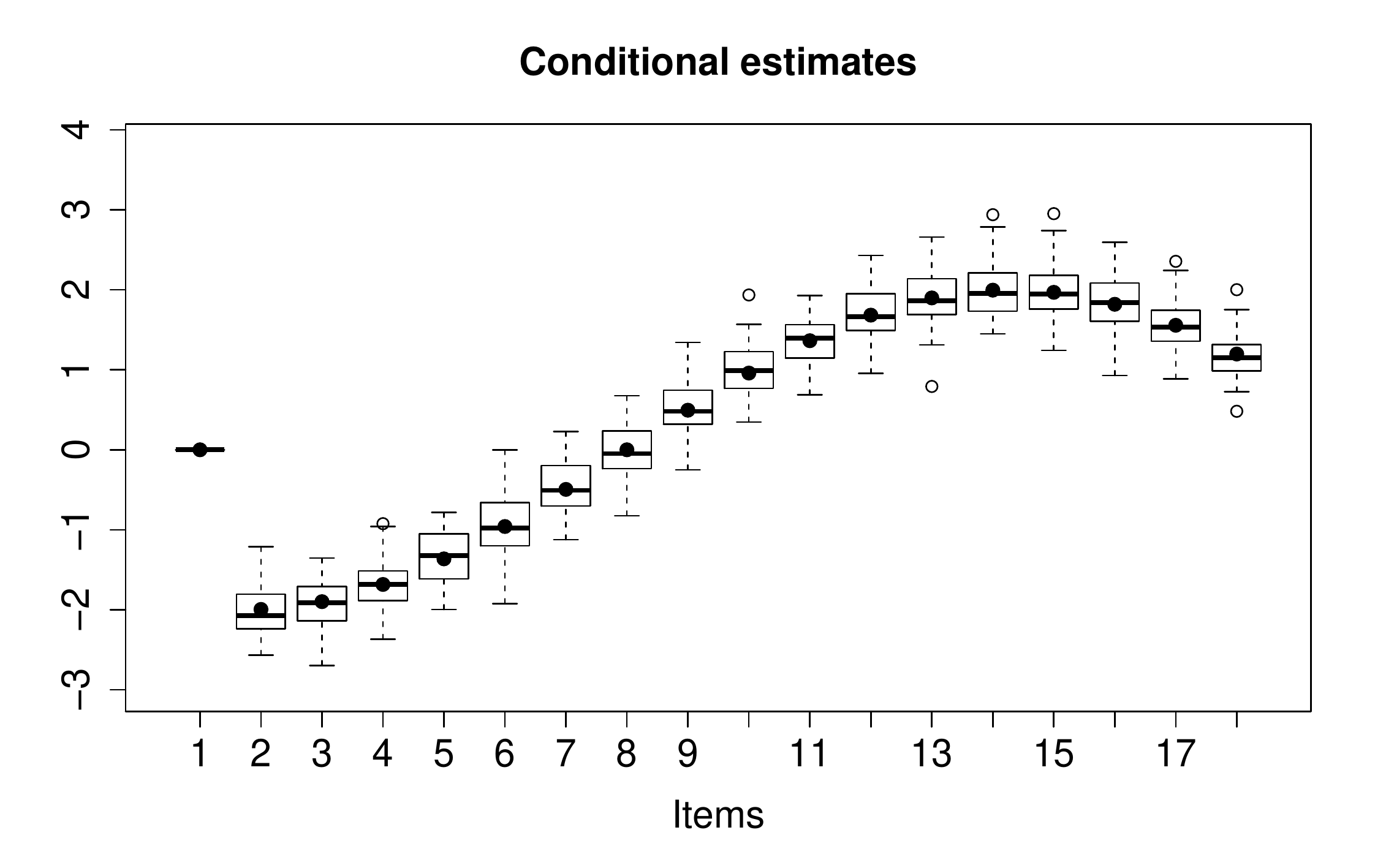}
\includegraphics[width=6.5 cm]{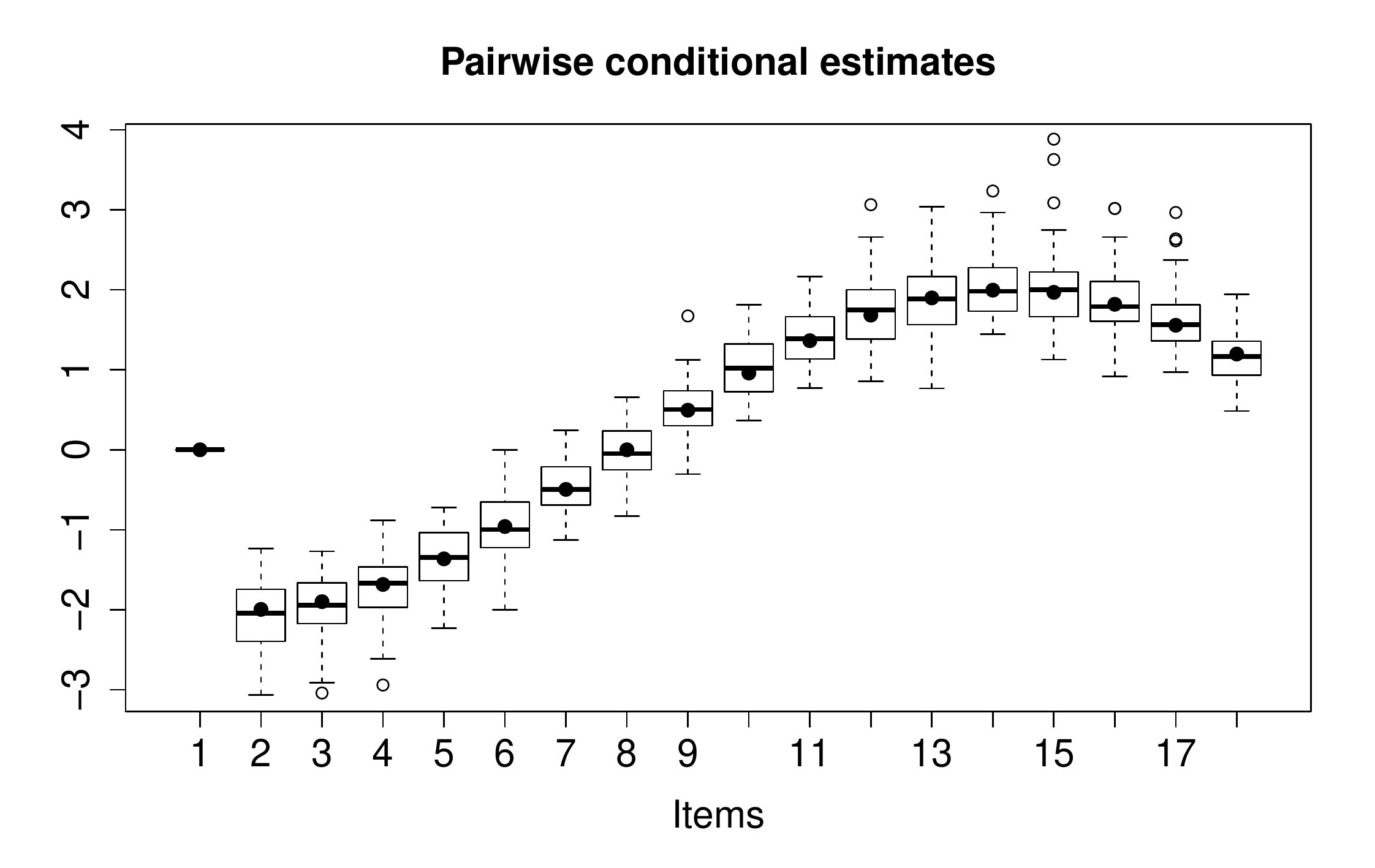}
\includegraphics[width=6.5 cm]{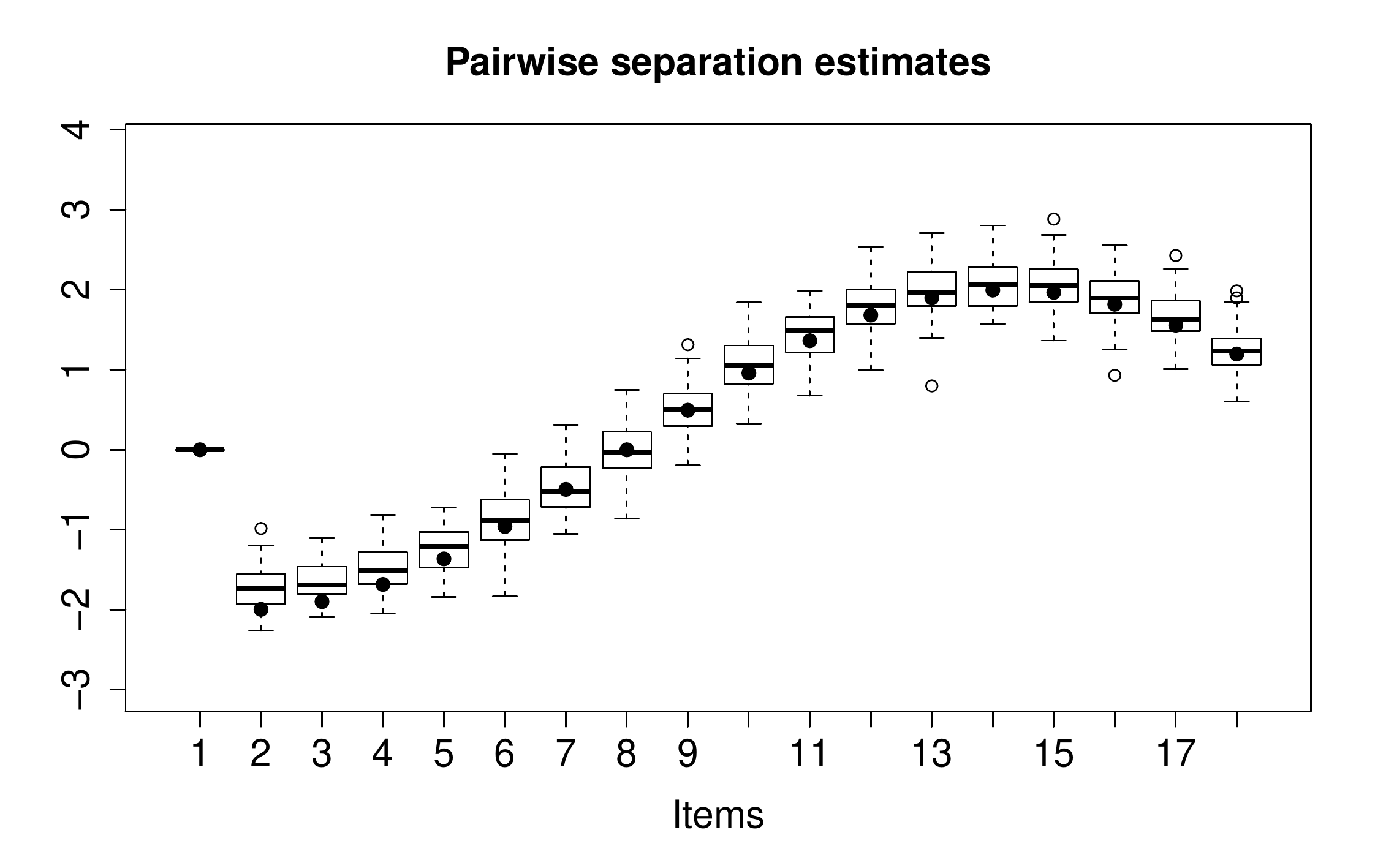}
\caption{Box plots for estimators; generating model is the binary Rasch model, persons drawn from $N(0,1)$ ($P=100, I=18$), dots indicate the true parameters. }
\label{fig:Raschvar100}
\end{figure}

Figure \ref{fig:Raschvar100} compares the performance of the pairwise estimators when persons are drawn from a $\chi$-squared distribution (binary Rasch model, persons drawn from $N(0,1)^2$, $P=100, I=18$). While both estimators show the same central tendency the conditional in some simulations the conditional pairwise estimator fits badly yielding estimates far from the true values. 

\begin{figure}[H]
\centering
\includegraphics[width=6.5 cm]{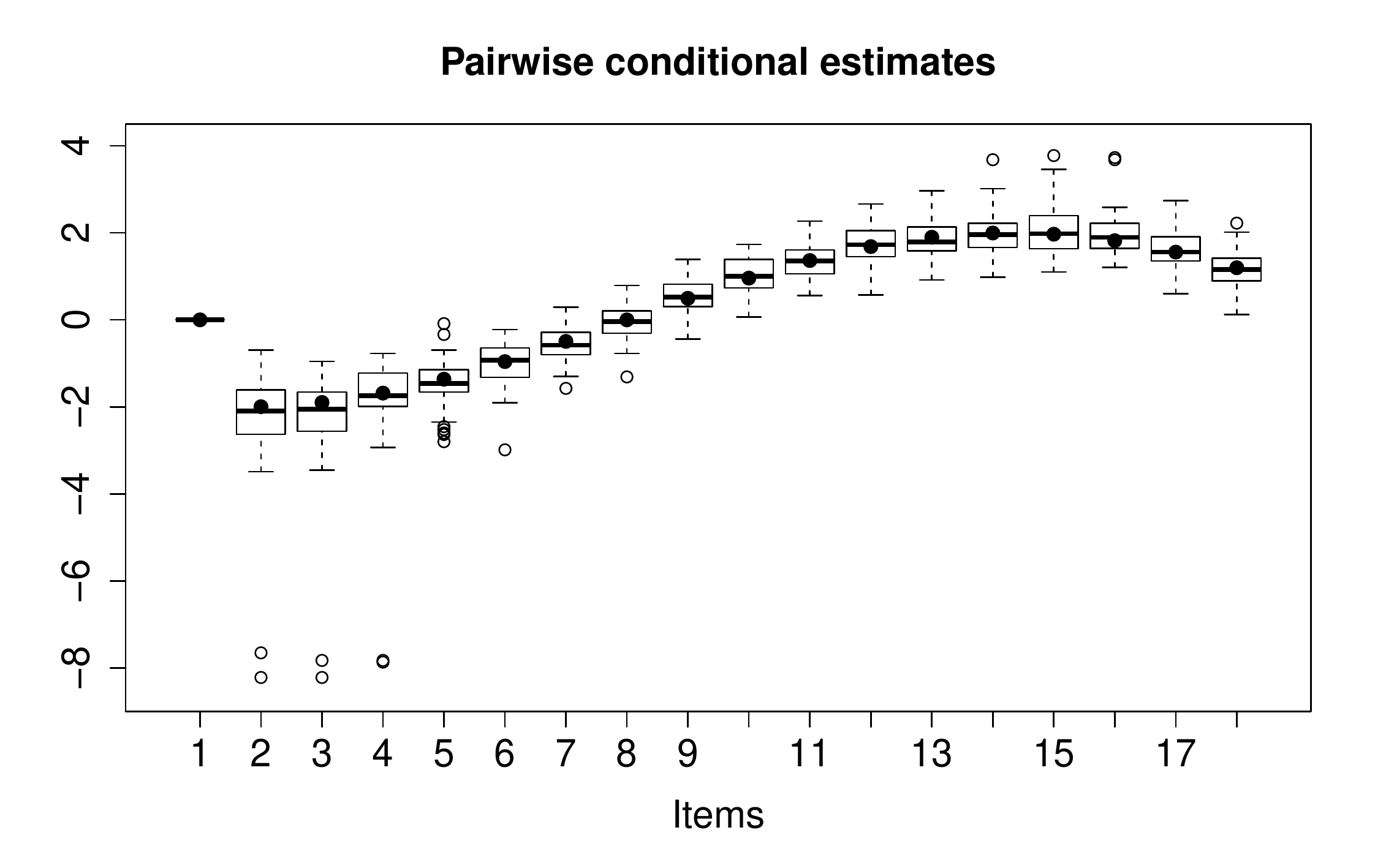}
\includegraphics[width=6.5 cm]{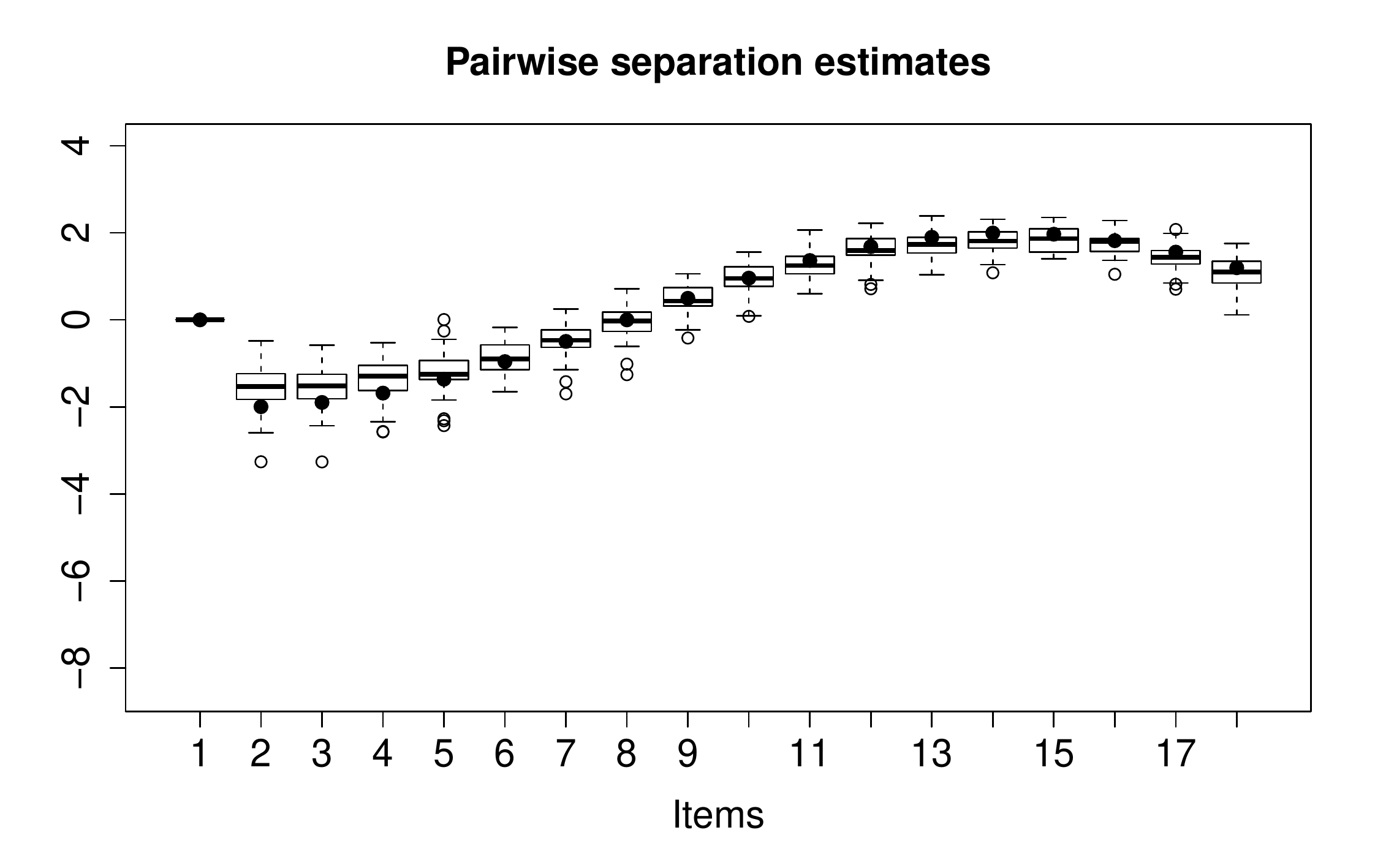}
\caption{Box plots for estimators; generating model is the binary Rasch model, persons drawn from $N(0,1)^2$ ($P=100, I=18$), dots indicate the true parameters. Conditional estimate did not exist in any of the simulations.}
\label{fig:Raschvar100}
\end{figure}

\end{document}